\documentclass[structabstract]{aa}  
\usepackage{graphicx}
\usepackage{txfonts}
\usepackage{aalongtable}
\usepackage{ulem}
\usepackage{natbib}
\bibpunct{(}{)}{;}{a}{}{,}
\begin{document}
   \title{New companions in the stellar systems of DI Cha, Sz 22, CHXR 32, and Cha H$\alpha$ 5 in the Cha I star-forming region
\thanks{Based on observations made with ESO telescopes at the Paranal Observatory 
under programme IDs 076.C-0292(A), 080.C-0424(A), 082.C-0489(A) and data obtained from the  ESO/ST-ECF Science Archive Facility from the Paranal Observatory under programme
ID 076.C-0579(A) and from the Hubble Space Telescope under programme ID SNAP-8216.}
}

  \titlerunning{New companions in the stellar systems of DI Cha, Sz 22, CHXR 32, and Cha H$\alpha$ 5 in Cha I}

   \author{T. O. B. Schmidt\inst{1}
           \and
           N. Vogt\inst{2}
           \and
           R. Neuh\"auser\inst{1}
           \and
           A. Bedalov\inst{1,3}
           \and
           T. Roell\inst{1}
}

   \offprints{Tobias Schmidt, e-mail:tobi@astro.uni-jena.de}

   \institute{Astrophysikalisches Institut und Universit\"ats-Sternwarte, Universit\"at Jena, Schillerg\"a\ss chen 2-3, 07745 Jena, Germany\\
              \email{tobi@astro.uni-jena.de}
         \and
          Departamento de F\'isica y Astronom\'ia, Universidad de Valpara\'iso, Avenida Gran Breta\~na 1111, Valpara\'iso, Chile
         \and
            Faculty of Natural Sciences, University of Split, Teslina 12. 21000 Split, Croatia
            }

   \date{Received 2012; accepted }

 
  \abstract
   {The star-forming regions in Chamaeleon (Cha) are among the nearest (distance $\sim$165 pc) and youngest (age $\sim$2 Myrs) conglomerates of recently formed stars and
among the ideal targets for studies of star formation.}
   {We search for new, hitherto unknown binary or multiple-star components and investigate their membership in Cha and their gravitationally bound nature.}
   {We used the Naos-Conica (NACO) instrument at the Very Large Telescope Unit 4/YEPUN of the Paranal Observatory, at 2 or 3 different epochs, in order to obtain relative and absolute
astrometric measurements, as well as differential photometry in the J, H, and Ks band. On the basis of known proper motions and these observations, we analysed the astrometric
results in our proper motion diagrams (PMD: angular separation / position angle versus time) to eliminate possible (non-moving) background stars
and establish co-moving binaries and multiples.}
   {DI Cha turns out to be a quadruple system with a hierachical structure, consisting of two binaries: a G2/M6 pair and a co-moving pair of two M5.5 dwarfs. For both pairs we
detected orbital
motion (P$\sim$130 and $\sim$65 years respectively), although in opposite directions. Sz 22 is a binary whose main component is embedded in a circumstellar disc or
reflection
nebula, accompanied by a co-moving M4.5 dwarf. CHXR 32 is a triple system, consisting of a single G5 star, weakened by an edge-on disc and a co-moving pair of M1/M3.5 dwarfs whose components show
significant variations in their angular separation. Finally, Cha H$\alpha$ 5 is a binary consisting of two unresolved M6.5 dwarfs whose strong variations in position angle at
its projected separation of only 8 AU imply an orbital period of $\sim$46 years. DI Cha D and Cha H$\alpha$ 5 A \& B are right at the stellar mass limit and could possibly be brown
dwarfs.}
   {In spite of various previously published studies of the star-forming regions in Cha we still found four hitherto unknown components in young low-mass binaries and multiple systems. All are
gravitationally bound, and at least the case Cha H$\alpha$ 5 presents a link between our high-resolution astrometry and the radial velocity method, avoiding
a blind gap of detection possibility. 
}

   \keywords{Stars: imaging, pre-main sequence, binaries: visual, brown dwarfs -- Astrometry -- Infrared: stars -- Protoplanetary discs}

   \maketitle
%

\section{Introduction}
\label{Section1}

\begin{table*}
\caption{Observed objects in Chamaeleon I}
\label{table:1}
\begin{center}
\begin{tabular}{lllcccccccc}
\hline\hline
Object$^a$            & RA           & Dec                & System           & Bin.  & Spectral        & SpT   & Dist.       & Dist. & 2MASS            & Backg.$^e$ \\
& [h \ m \ s]$^{a,b}$ & [$^{\circ}$ \ \ $'$ \ \ $''$]$^{a,b}$ & architecture$^c$ & Ref.  & types           & Ref.  & [pc]        & Ref.  & K\,[mag]$^d$     & density    \\
\hline
DI Cha                & 11 07 20.72  & -77 38 07.3        & **+**            & 1,2,3 & G2+M6+M5.5+M5.5 & 4,3   & \ \ 223$^f$ & 5     & \ 6.217          & 0.070      \\
Sz 22                 & 11 07 57.93  & -77 38 44.9        & **+w**           & 3,6,7 & K7+M4.5+M2+M3   & 8,3,9 & 165         & Cha I & \ 6.830          & 0.070      \\
CHXR 32               & 11 08 14.94  & -77 33 52.2        & *+**             & 10,3  & G5+M1+M3.5      & 11,3  & 165         & Cha I & \ \ \ 6.182$^g$  & 0.070      \\
Cha H$\alpha$ 5       & 11 08 24.11  & -77 41 47.4        & **               & 3     & M6.5+M6.5       & 12,3  & 165         & Cha I & 10.711           & 0.091      \\
\hline
\end{tabular}
\end{center}
\textbf{Notes.} $^{(a)}$ Taken from the SIMBAD database \citep{2007ASPC..377..197W}. $^{(b)}$ 
International Celestial Reference System (ICRS) coordinates (epoch=J2000). $^{(c)}$ Updated multiplicity
of the objects: *: star, **: binary, w**: wide binary stellar companion candidate, please see text for previously known multiplicity.  
$^{(d)}$ \citet{2006AJ....131.1163S,2003tmc..book.....C}. $^{(e)}$ expected number of background stars in the NACO 
S13 field of view (see text). $^{(f)}$ see text for a discussion. $^{(g)}$ value by \citet{2002AJ....124.1001C}, as 2MASS only lists an upper limit, according to non-detection. \\
\textbf{References.} (1) \citet{1993A&A...278...81R}. (2) Simultaneously imaged by \citet{2008ApJ...683..844L} and by us (here).
(3) newly found (here). (4) \citet{1973ApJ...180..115H}. (5) \citet{2007A&A...474..653V}. (6) \citet{1997ApJ...481..378G}. (7) \citet{2008ApJ...683..844L}.
(8) \citet{1999A&A...343..477C}. (9) \citet{2003AJ....125.2134G}. (10) \citet{1988A&A...207...46C}. (11) \citet{1989ApJ...338..262F}. (12) \citet{2004A&A...416..555L}.
\end{table*}

The star-forming region Cha I \citep{2008hsf2.book..169L} is rather close to the Sun (distance $\sim$165 pc) and is young (age $\sim$2 Myrs), among the ideal targets for
population studies of young
stars, brown dwarfs, and planetary bodies. In this context, the occurrence and properties of binaries and multiple systems can play a key role in understanding 
the underlying physical
processes in the formation and evolution of recently formed stars and their companions. Our working group is carrying out a long-term programme, obtaining astrometry and
differential
photometry in the near-infrared JHKs bands from images at the Very Large Telescope (VLT at Paranal Observatory of the European Southern Observatory, ESO), Unit Telescope (UT) 4 (Yepun)
with
Naos-Conica \citep[NACO,][]{2003SPIE.4841..944L, 2003SPIE.4839..140R} at different epochs.  First results of this campaign have been published by \citet{2008A&A...491..311S,2008A&A...484..413S}
and by \citet[][hereafter Paper I]{2012A&A...546A..63V}.
The latter paper confirms 16 gravitationally bound binaries or multiple systems in the star-forming regions in Cha. The purpose of the
present paper is to complement this information, presenting and confirming new faint companions in a total of four stellar systems, all members of the star-forming region Cha I. 

In Section \ref{Section2} we describe our observations, as well as the background star density and its relation to the interstellar extinction in the surroundings of our targets.
In Section \ref{Section3} we present the individual target stars in detail, based on their images reproduced from our NACO fields and the corresponding proper motion diagrams
(PMDs). Section \ref{Section4} contains concluding remarks. 

\section{Observations and background star density}
\label{Section2}

For a detailed description of our observing and reduction strategy, and additionally for the calibration procedure and the astrometric analysis using the 
proper motion diagram (PMD), we refer to
Paper I. Table \ref{table:1} contains the main properties of our four target stars and the related references, while in Table \ref{table:2} we list our VLT/NACO observation log. 
Table \ref{table:3} contains the astrometric calibration results and Table \ref{table:4} the proper motion values used. The absolute and relative astrometric results are listed in Tables
\ref{table:5} and \ref{table:6}, respectively, while the resulting gradients of the variations in angular separations and in position angle are given in Tables \ref{table:7} and
\ref{table:8}, respectively, based on linear fits of the corresponding astrometric results.
In Table \ref{table:9} we summarize the differential photometry. In Figures 1 -- 8 we present detailed maps and the proper motion diagrams (PMDs) for each of the stellar systems. 

\begin{table*}
\caption{VLT/NACO observation log}
\label{table:2}
\begin{center}
\begin{tabular}{lllcccccccccc}
\hline\hline
Object           & Other name            & JD - 2453700      & Date of         & DIT        & NDIT   & Number    & Airmass & DIMM$^b$& $\tau_{0}^c$ & Filter \\
                 &                       & $[\mathrm{days}]$ & observation     & [s]        &        & of images &         & Seeing  & [ms]         &        \\
\hline
\object{DI Cha}  & HIP 54365             & \ \ \ \ 83.85826  & 17 Feb 2006     & 0.3454     & 100    & 20        & 1.79    & 0.63   & 6.5          & Ks     \\
                 &                       & \ \ 815.62668     & 19 Feb 2008     & 0.3454     & 174    & 5         & 1.82    & 0.80   & 5.3          & Ks     \\
                 &                       & \ \ 815.63134     & 19 Feb 2008     & 1.2        & 50     & 5         & 1.81    & 0.84   & 4.7          & J      \\
                 &                       & 1182.64078$^a$    & 20 Feb 2009     & 0.3454     & 87     & 20        & 1.77    & 0.81   & 5.1          & Ks     \\
\object{Sz 22}   & FK Cha                & \ \ \ \ 83.88427  & 17 Feb 2006     & 3          & 12     & 20        & 1.86    & 0.56   & 6.9          & Ks     \\
                 &                       & \ \ 815.64511     & 19 Feb 2008     & 4          & 15     & 5         & 1.77    & 0.86   & 4.7          & Ks     \\
                 &                       & \ \ 815.65010     & 19 Feb 2008     & 30         & 2      & 5         & 1.76    & 0.80   & 5.0          & J      \\
\object{CHXR 32} & Glass I / HP Cha      & \ \ 816.89621     & 20 Feb 2008     & 1/3        & 60/20  & 5/5       & 1.92    & 1.65   & 2.3          & Ks     \\
                 &                       & 1182.90637$^a$    & 20 Feb 2009     & 0.3454/0.5 & 87/120 & 12/6      & 1.97    & 0.52   & 8.9          & Ks     \\
Cha H$\alpha$ 5  & \object{ISO-ChaI 144} & \ \ \ \ 82.75127  & 16 Feb 2006     & 50         & 1      & 20        & 1.66    & 0.55   & 8.3          & Ks     \\
                 &                       & \ \ 816.68649     & 20 Feb 2008     & 60         & 1      & 5         & 1.70    & 0.79   & 4.8          & Ks     \\
                 &                       & \ \ 816.69139     & 20 Feb 2008     & 60         & 1      & 5         & 1.69    & 0.71   & 5.4          & J      \\
                 &                       & 1181.73805$^a$    & 19 Feb 2009     & 30         & 2      & 15        & 1.66    & 0.42   & 7.5          & Ks     \\
\hline
\end{tabular}
\end{center}
\textbf{Notes.} Each image consists of the number of exposures given in column 6 times the individual integration 
time given in column 5. $^{(a)}$ Data taken in cube mode, so each image is a cube of the number of planes given in column 
6, each having the individual integration time given in column 5. $^{(b)}$ Differential image motion monitor (DIMM) seeing average of all images taken from the individual fits
headers. $^{(c)}$ coherence time of atmospheric fluctuations.
\end{table*}

\begin{table}
\caption{Astrometric calibration results using the binary \object{HIP 73357}}
\label{table:3}
\begin{center}
\begin{tabular}{lcccc}
\hline\hline
JD - 2453700      & Epoch        & Pixel scale & Orientation & Filter \\
$[\mathrm{days}]$ &              & [mas/Pixel] & [$\degr$]       &        \\
\hline
\ \ \ \ 88.84779  & Feb 2006     & 13.24 $\pm$ 0.18 & 0.18 $\pm$ 1.24 & Ks \\
\ \ 815.91117     & Feb 2008     & 13.22 $\pm$ 0.20 & 0.73 $\pm$ 1.40 & J  \\
\ \ 815.91907     & Feb 2008     & 13.25 $\pm$ 0.20 & 0.68 $\pm$ 1.40 & Ks \\
1181.89936        & Feb 2009     & 13.25 $\pm$ 0.21 & 0.76 $\pm$ 1.48 & Ks \\
\hline
\end{tabular}
\end{center}
\textbf{Notes.} Hipparcos values \citep{1997A&A...323L..49P} were used as reference values. Measurement 
errors of Hipparcos as well as maximum possible orbital motion since the epoch of the Hipparcos observation, are 
taken into account.
\end{table}

\begin{table}
\caption{Proper motions}
\label{table:4}
\begin{center}
\begin{tabular}{llr@{\,$\pm$\,}lr@{\,$\pm$\,}l}
\hline\hline
Object & Reference             &\multicolumn{2}{c}{$\mu_{\alpha} \cos{\delta}$}    & \multicolumn{2}{c}{$\mu_{\delta}$}  \\
       &                       &\multicolumn{2}{c}{[mas/yr]}                       & \multicolumn{2}{c}{[mas/yr]}\\
\hline
DI Cha          & Hipparcos \\
                & (new) (1)            & -24.61  & 1.84                           & 3.45     & 1.54 \\

                & Tycho-2 (2)          & -23.6   & 3.0                            & 6.0      & 2.8  \\
\hline
                & used:  \\
                & UCAC 3 (3)           & -15.8   & 1.6                            & -5.1     & 1.7  \\
\hline
Sz 22           & UCAC 3 (3)           & -26.7   & 5.7                            & 13.7     & 7.7  \\
\hline
CHXR 32         & UCAC 3 (3)           & -2.1    & 5.6                            & -4.4     & 5.5  \\
                & PPMX (4)             & -27.2   & 3.8                            & 14.14    & 3.8  \\
\hline
                & used:  \\
                & UCAC 2 (5)           & -15.9   & 3.7                            & 7.7      & 3.4  \\
\hline
Cha H$\alpha$ 5 & PSSPMC (6)           & -18     & 13                             & 11       & 13   \\
                & SSS-FORS1 (7)        & -31.8   & 21.7                           & 12.5     & 21.7 \\
                & SSS-SofI (7)         & -29.6   & 11.8                           & 9.7      & 11.8 \\
\hline
                & weighted mean        & -25.4   & 8.1                            & 10.6     & 8.1  \\
\hline
Median Cha I    & Luhman$^a$ (8)       & -21    & $\sim$1                         & 2     & $\sim$1 \\
\hline
\end{tabular}
\end{center}
\textbf{Notes.} Only independent sources with individual error bars for the targets were considered. 
$^{(a)}$ Based on UCAC2 proper motions from (5) \\
\textbf{References.}
(1) \citet{2007A&A...474..653V} (2) \citet{2000A&A...355L..27H} (3) \citet{2010AJ....139.2184Z}
(4) \citet{2008A&A...488..401R} (5) \citet{2004AJ....127.3043Z} (6) \citet{2005A&A...438..769D} 
(7) own data (here) $\&$ \citet{2001MNRAS.326.1279H} (8) \citet{2008ApJ...675.1375L}
\end{table}

The last column of Table 1 refers to the expected number of fore- and background stars in the NACO S13 field (field of view 13.56 x 13.56 arcseconds), according to the star counts down to
the 2MASS limiting magnitude (near K = 16 mag) in a cone of a radius of 300 arcseconds around each target. The mean density value of our four targets is 0.075 $\pm$ 0.009,
which is consistent with the
average value 0.118  $\pm$ 0.030 of the 16 binaries and multiple systems in Cha, described in Paper I. Probably, the slightly lower background star number density by 1.4 $\sigma$ here
occurs because these four stars are embedded in the central cloud of Cha I, with a stronger interstellar extinction A$_V$ of 6.8 $\pm$
1.9 mag according to an extinction map by 
\citet{2006A&A...447..597K} or 4.6 $\pm$ 0.8 mag according to an extinction map by \citet{1997A&A...324L...5C}, while most targets in Paper I are distributed over a much larger area on the
sky. This extinction value is in both cases a factor of 1.3 $\pm$ 0.4 higher than for all of the 16 multiple systems from Paper I within the region of the extinction maps, only
weakly supporting the
above-mentioned explanation, while a strong argument in favour of this hypothesis is that the objects SZ Cha, RX J1109.4-7627, and Sz 41 from Paper I, all being slightly
outside the central region, possess only about 1.2 mag extinction in contrast to the eight most central objects exhibiting
on average about 4.3 mag of extinction \citep{1997A&A...324L...5C}.


\section{Description of the individual stars}
\label{Section3}

\subsection{The quadruple system  DI Cha}

\begin{figure*}
\centering
\resizebox{0.75\textwidth}{!}{\includegraphics{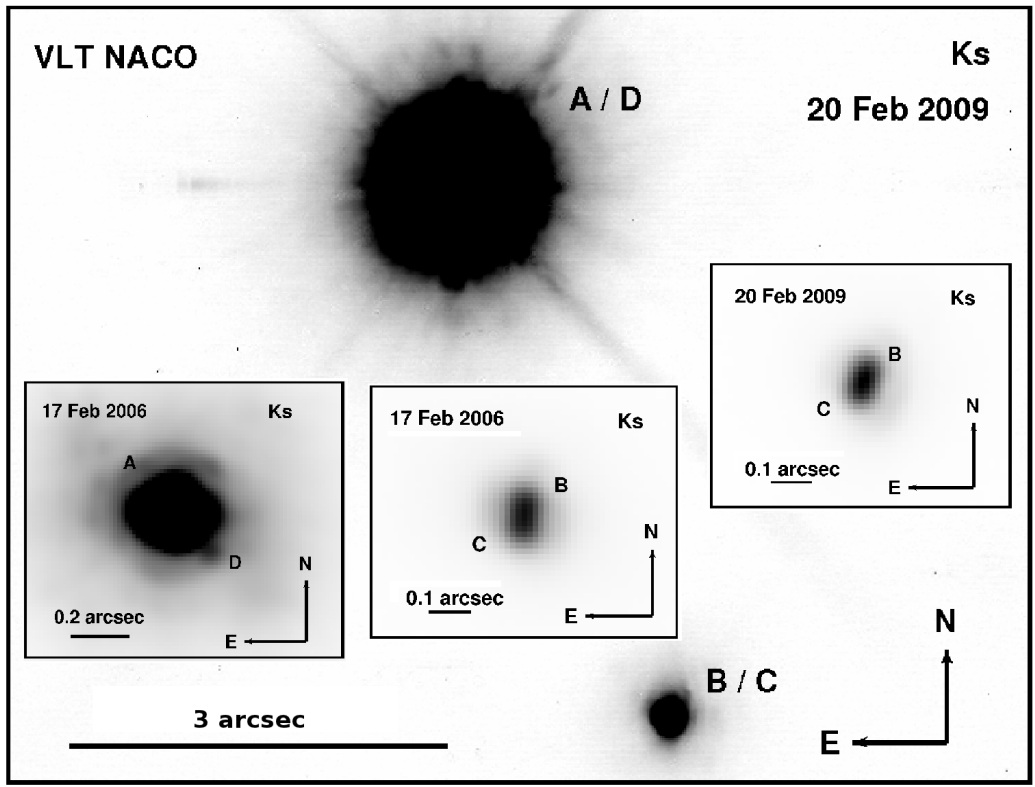}}
\caption{VLT NACO Ks band images of DI Cha. Main frame: One of our observations of the first found binarity of DI Cha \citep{1993A&A...278...81R}. \textit{Inserts}: our new detected component
D (left insert), and two images of the B/C pair at different epochs, revealing orbital motion in the position angle (central and right insert).}
\label{FigDICha}
\end{figure*}

\begin{figure*}
   \centering
   \includegraphics[width=0.39\textwidth]{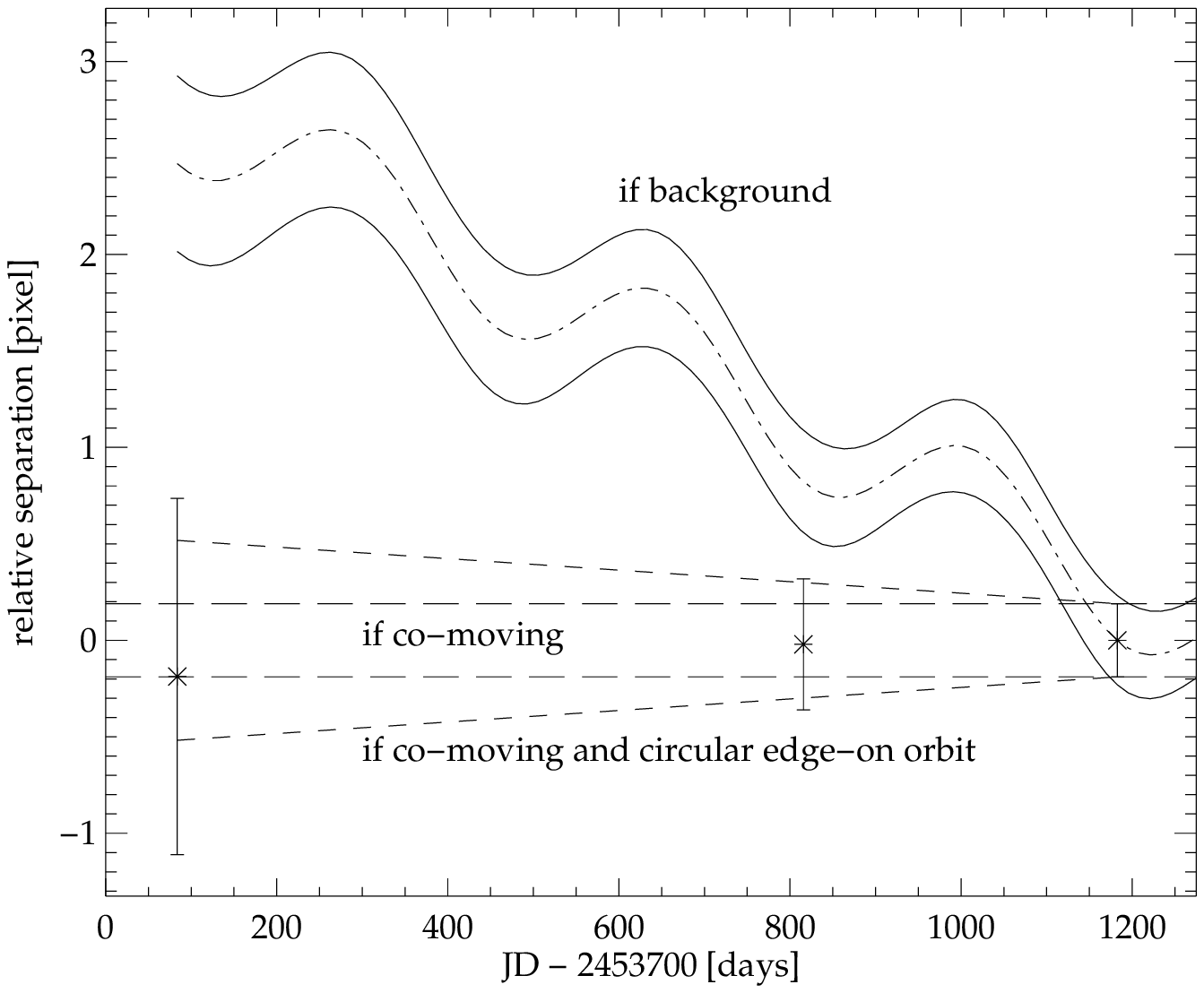}
   \includegraphics[width=0.39\textwidth]{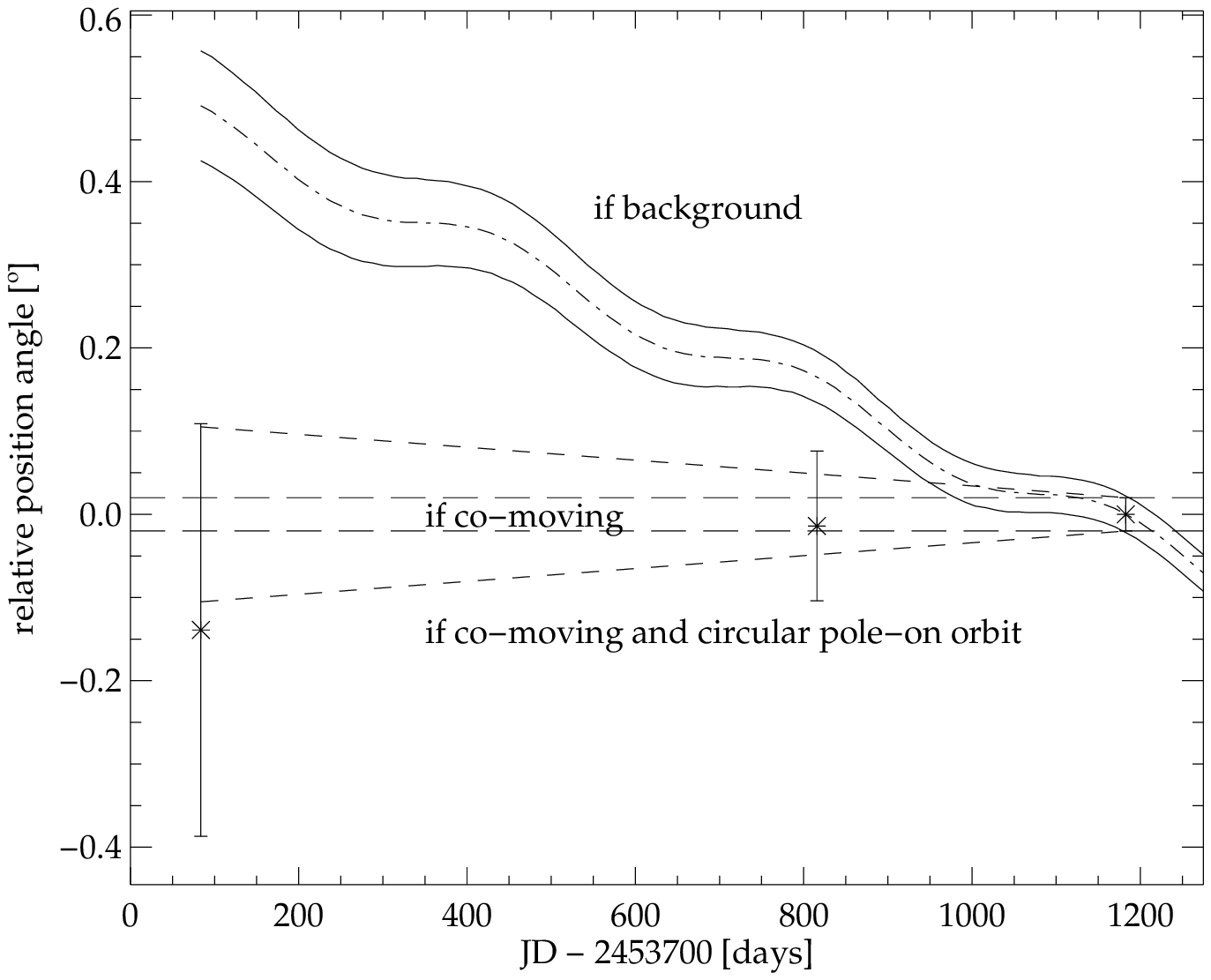}
   \includegraphics[width=0.39\textwidth]{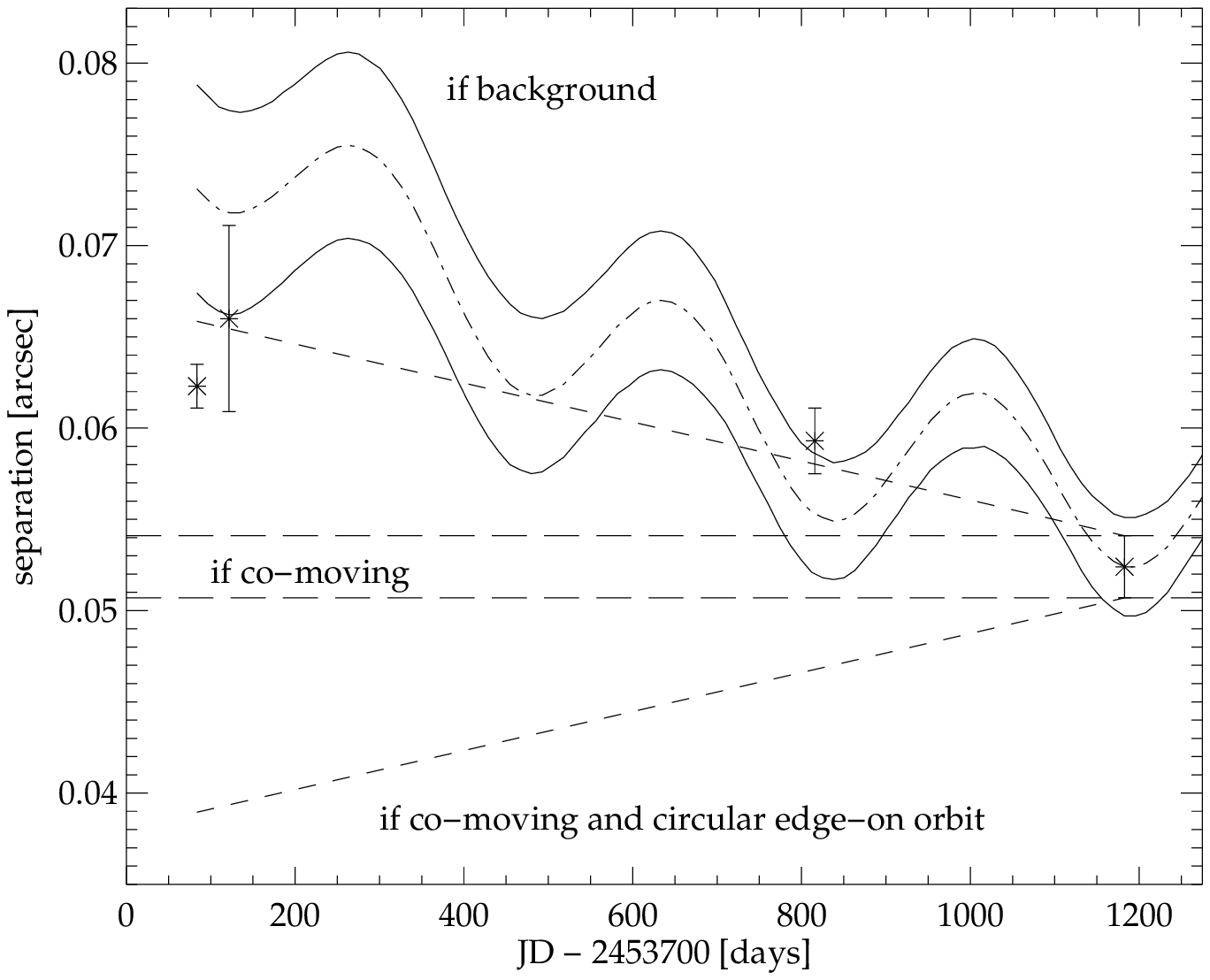}
   \includegraphics[width=0.39\textwidth]{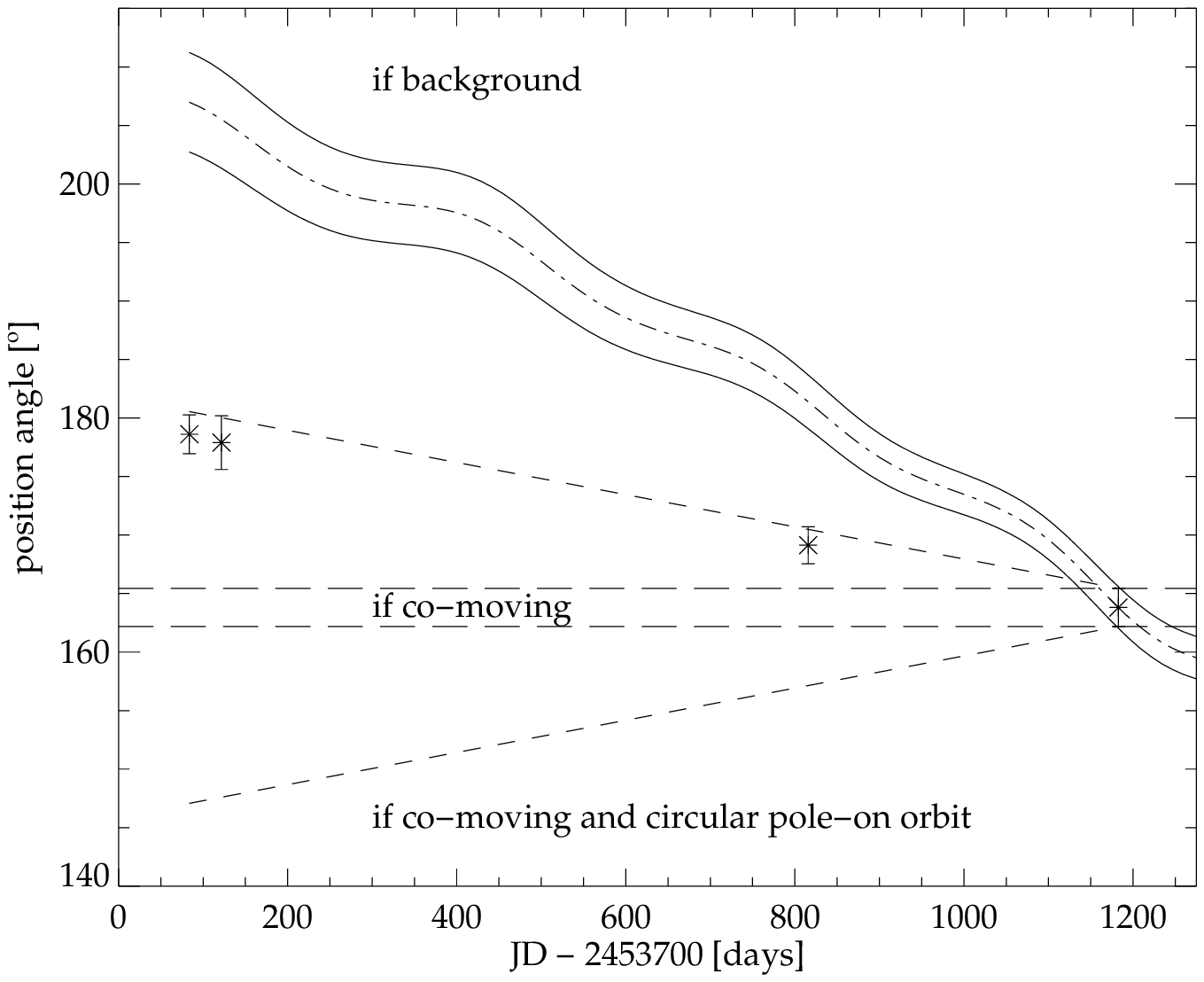}
   \includegraphics[width=0.39\textwidth]{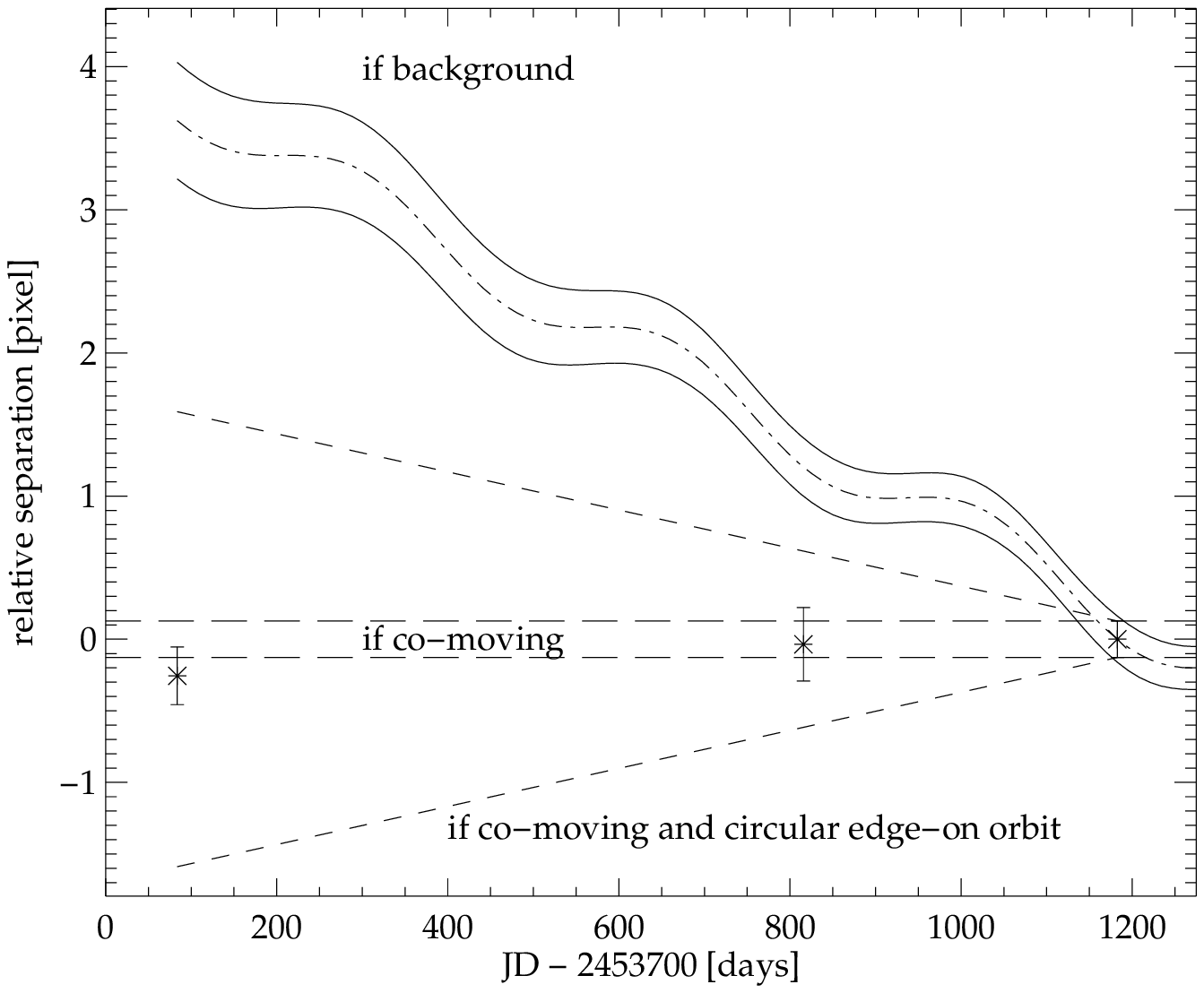}
   \includegraphics[width=0.39\textwidth]{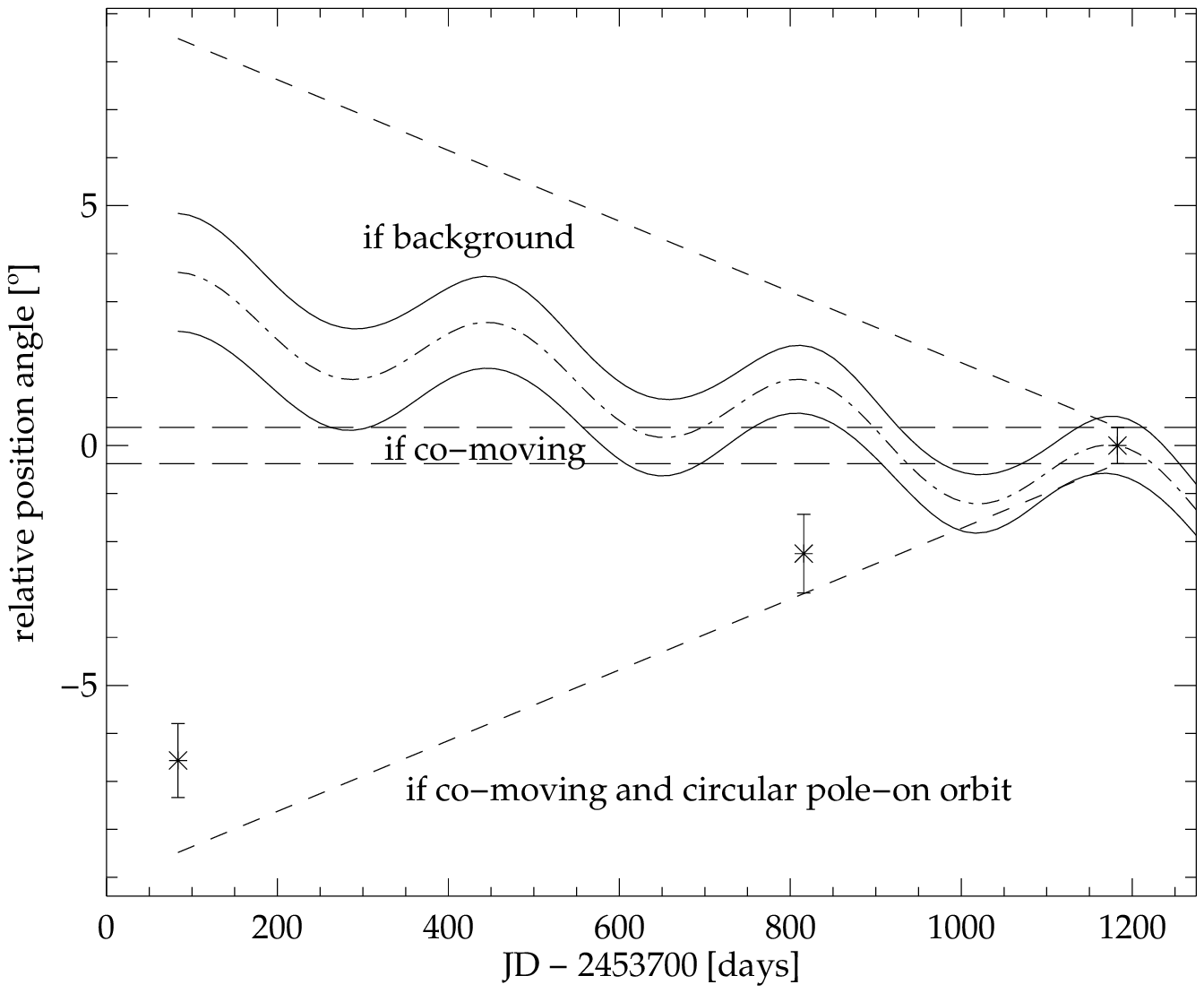}
   \caption{Proper motion diagrams (PMD) of different components in the DI Cha quadruple system for separation (left) and position angle (right). \textit{Top:}
relative astrometric measurements of the centre of gravity of B/C (assuming being approx.~the centre of light) relative to A. \textit{Middle:} absolute astrometric measurements of the pair B/C.
\textit{Bottom:} relative astrometric measurements of D relative to A.}
   \label{FigDIChaPMD}
\end{figure*}

According to \citet{2007A&A...467.1147G} and \citet{2003A&A...410..269M}, DI Cha is spectroscopically a single star with a relatively high mass of $\sim$2.4 
M$_{\sun}$ and spectral type around G2 \citep{2007A&A...462..211L}. According to \citet{2011MNRAS.410..190T}, the star has an age of 2.7\,$\pm$\,1.8 Myr and a mass of 2.0\,$\pm$\,0.1
M$_{\sun}$. The Paschen $\beta$ line appears in emission \citep{2003AJ....125.2134G}, and \citet{2011ApJS..193...11M} report that DI Cha is a Class II
object with mid-IR excess emission arising from a protoplanetary disc surrounding it, but with relatively low mass
\citep[$\leq$ 0.009 M$_{\sun}$,][]{2007A&A...462..211L}. \citet{1993A&A...278...81R} detected a faint
companion candidate at an angular separation
$\sim$4.6'', later resolved into a binary source itself by \citet{2008ApJ...683..844L}, and separated by only $\sim$0.06'' or 10 AU,
denoted here as B and C. Here we present slightly earlier data of this triplicity from February
2006, as well as the first confirmation of the triple system, based on common proper motion, and not solely based on statistical arguments.

In addition to these three known components, our observations reveal another companion candidate at a separation of 0.2'' from DI Cha A (Fig. \ref{FigDICha}), 
designated D. Our PMDs 
(Fig. \ref{FigDIChaPMD}) exclude the background hypothesis, and all four stars are co-moving. However, there is no significant orbital motion of the fainter pair B/C around the primary
pair A/D. On the other hand, both close pairs reveal significant orbital variations, especially of their position angles. That of pair B/C is compatible
with an orbital period of about 65 years, that of A/D with about 130 years. It seems remarkable that this orbital motion seems to be clockwise in the case of B/C, but counter-clockwise in A/D.
From the middle panels in Fig. \ref{FigDIChaPMD} we find that because both the change in PA and separation are 
between the expectation for edge-on and face-on orbits, we can conclude that the orbit inclination of DI Cha B \& C lies between
these two extremes or that the eccentricity is high. Likewise we find from the bottom panels that the (non-detection
of a) change in separation, and that the change in PA is close to the expectation for a face-on orbit, we can conclude that the orbit of DI Cha
D around the primary is face-on - and/or that the eccentricity is high.
These results mean that co-planarity of the orbits of DI Cha D around the primary and DI Cha C around B is impossible. 

According to their magnitudes (Table \ref{table:9}), all three fainter components should be M type dwarfs. Although the magnitude of DI Cha D is
similar to that of Cha H$\alpha$ 2 B, which is a brown dwarf candidate  \citep{2008A&A...484..413S}, the larger parallactic distance to the DI Cha system of 223\,$\pm$\,79 pc
\citep{2007A&A...474..653V}
diminishes the probability of a brown dwarf classification of DI Cha D. In \citet{1999A&A...352..574B} the authors confirm
that YSOs are located in their associated clouds, as anticipated by a large body of work, and discuss reasons that make the individual parallaxes of some YSOs doubtful.
They discuss that Chamaeleon I is at the previously anticipated distance and finally find DI Cha to be at 194 $\pm$ 58 pc if single.
We therefore assume that DI Cha is at the distance of Cha I, with a best
distance estimate of 165 $\pm$ 30 pc as discussed in \citet{2008A&A...491..311S}.

On the basis of these previous assumptions we used the COND evolutionary models \citep{2003A&A...402..701B}, the BCAH models
\citep{1998A&A...337..403B,2002A&A...382..563B}, and
the temperature (to spectral type conversion) scale of \cite{2003ApJ...593.1093L} to estimate a spectral type of M6 $\pm$ 1.5 for the new companion. 
While the same models give a best mass estimate above the lower stellar limit, the lower 1\,$\sigma$ mass errors point to a minimum mass of $\geq$ 67
M$_{\mathrm{Jup}}$ for DI Cha D, justifying a brown dwarf candidate classification.

The same procedure results in a spectral type of M5.5 $\pm$ 1.5 for the components B and C, just above the brown dwarf -- stellar mass boundary and much later than the spectral type K0 found
by \citet{2007ApJ...662..413K}. Common for all three low-mass companions DI Cha B -- D is that their J-Ks colour is too red for their estimated spectral types (Table \ref{table:9}).

While we could already prove that DI Cha D is 82\,\% likely to be a real companion candidate, inspecting the speckle
pattern of our individual measurements in the first epoch with a new code\footnote{available at http://www.cran.r-project.org/web/packages/ringscale/}
developed in R by \citet{HaaseD2009}, we can confirm this result by using all three epochs of DI Cha D, showing a linear counter-clockwise orbital motion (Fig.~\ref{FigDIChaPMD},
Tables \ref{table:7}\,\&\,\ref{table:8}).

\subsection{The binary Sz 22}

\begin{figure*}
\centering
\resizebox{0.37\textwidth}{!}{\includegraphics{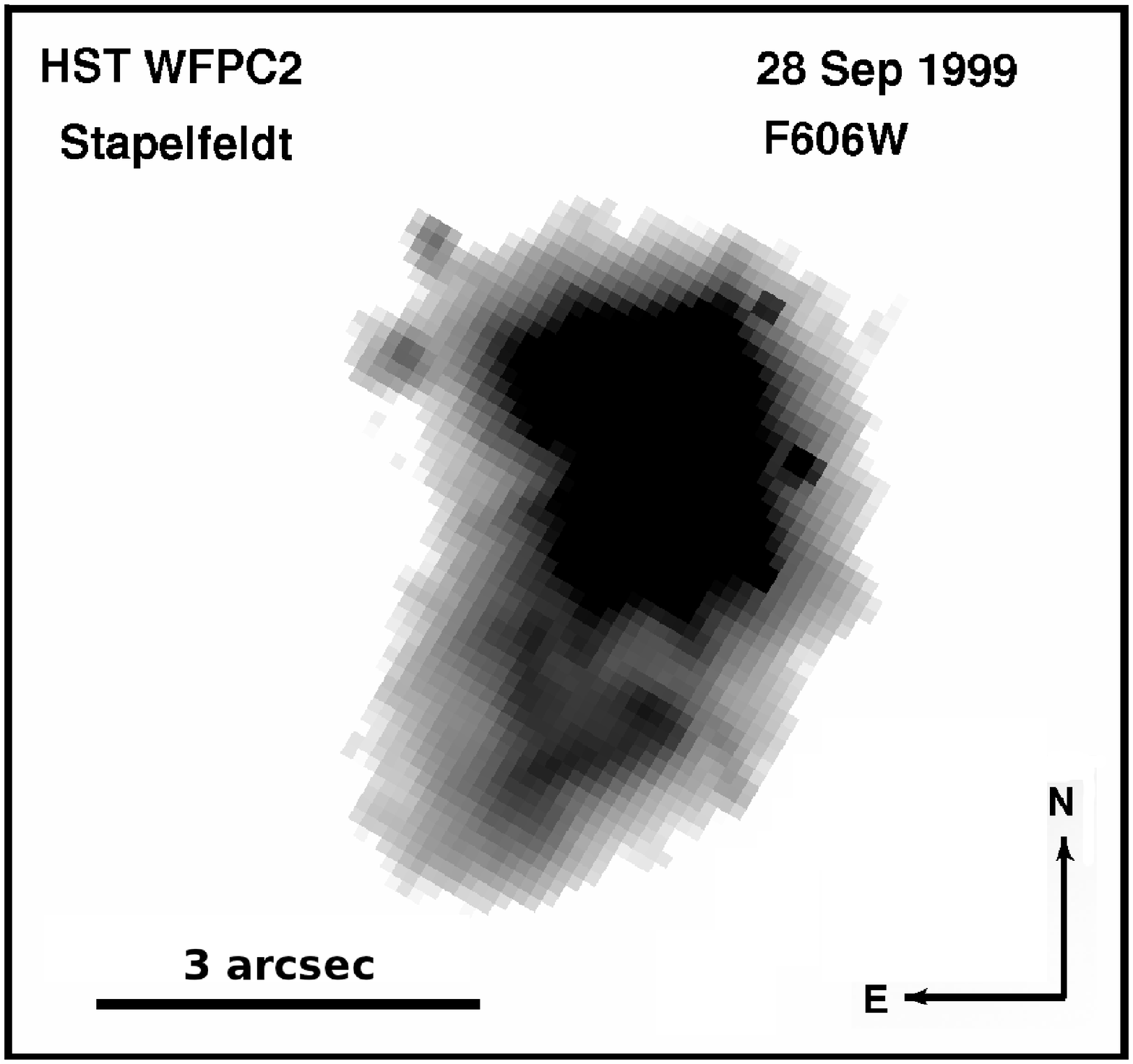}}
\resizebox{0.485\textwidth}{!}{\includegraphics{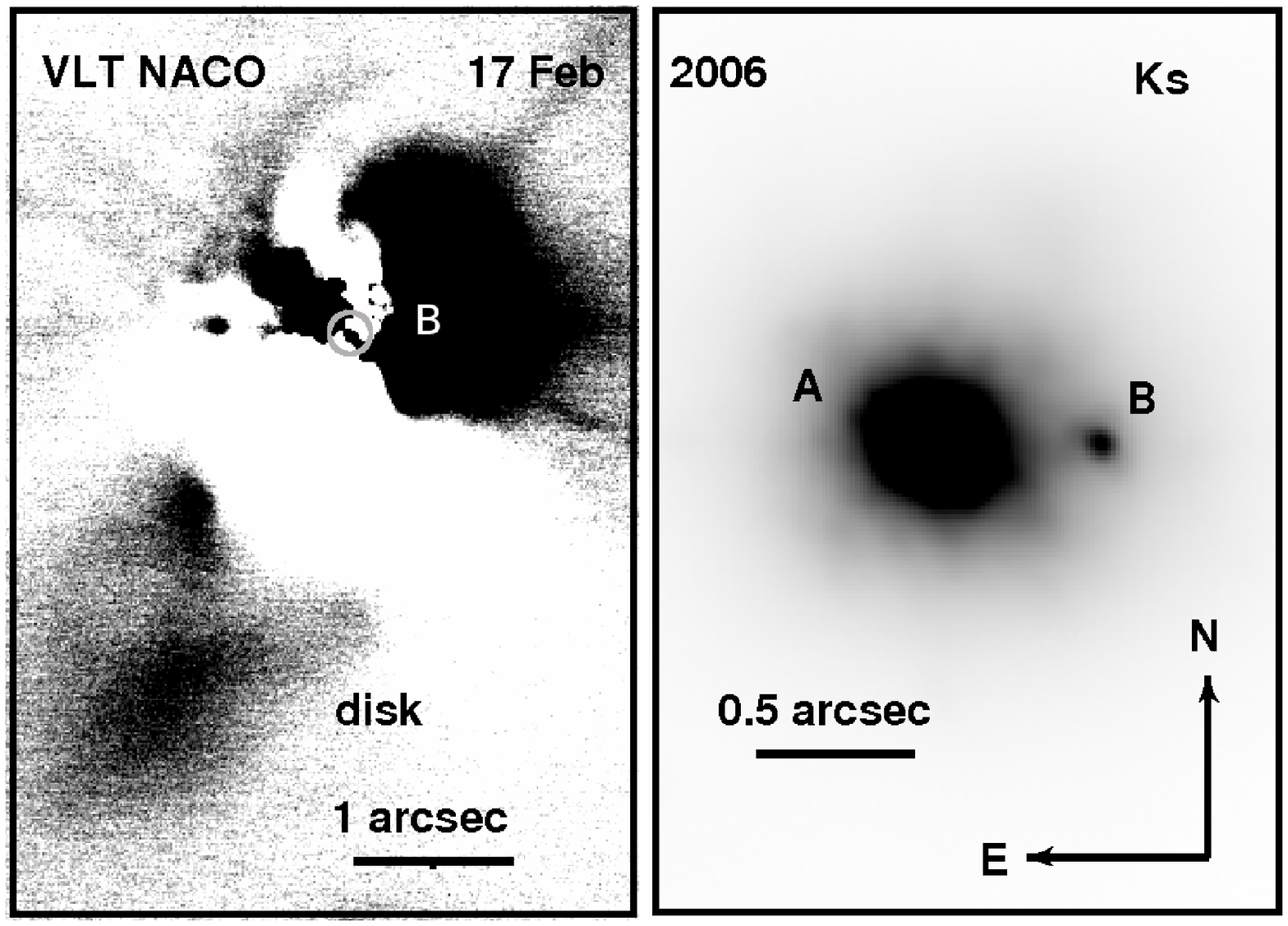}}
\caption{Sz 22 and its surroundings:  Images of the circumstellar disc/reflection nebula in the visual band pass (HST, left) and in the infrared (VLT NACO Ks band, centre). The right frame
shows Sz 22 with its newly detected stellar companion B (same image as in the centre, but without PSF subtraction of the primary positioned at the light grey circle in
the central image). Please see text for further information.}
\label{FigSZ22}
\end{figure*}

\begin{figure*}
   \centering
   \includegraphics[width=0.39\textwidth]{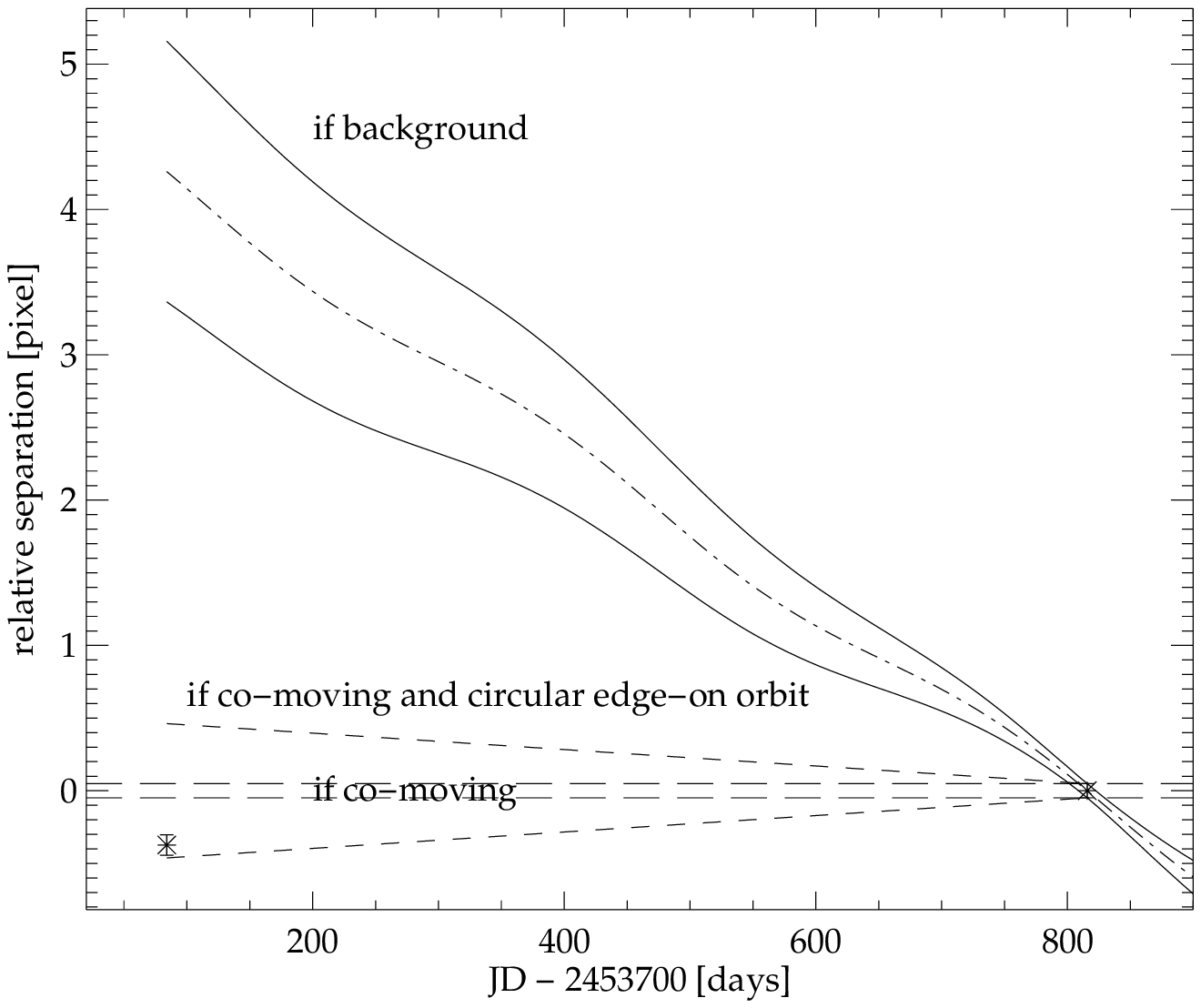}
   \includegraphics[width=0.39\textwidth]{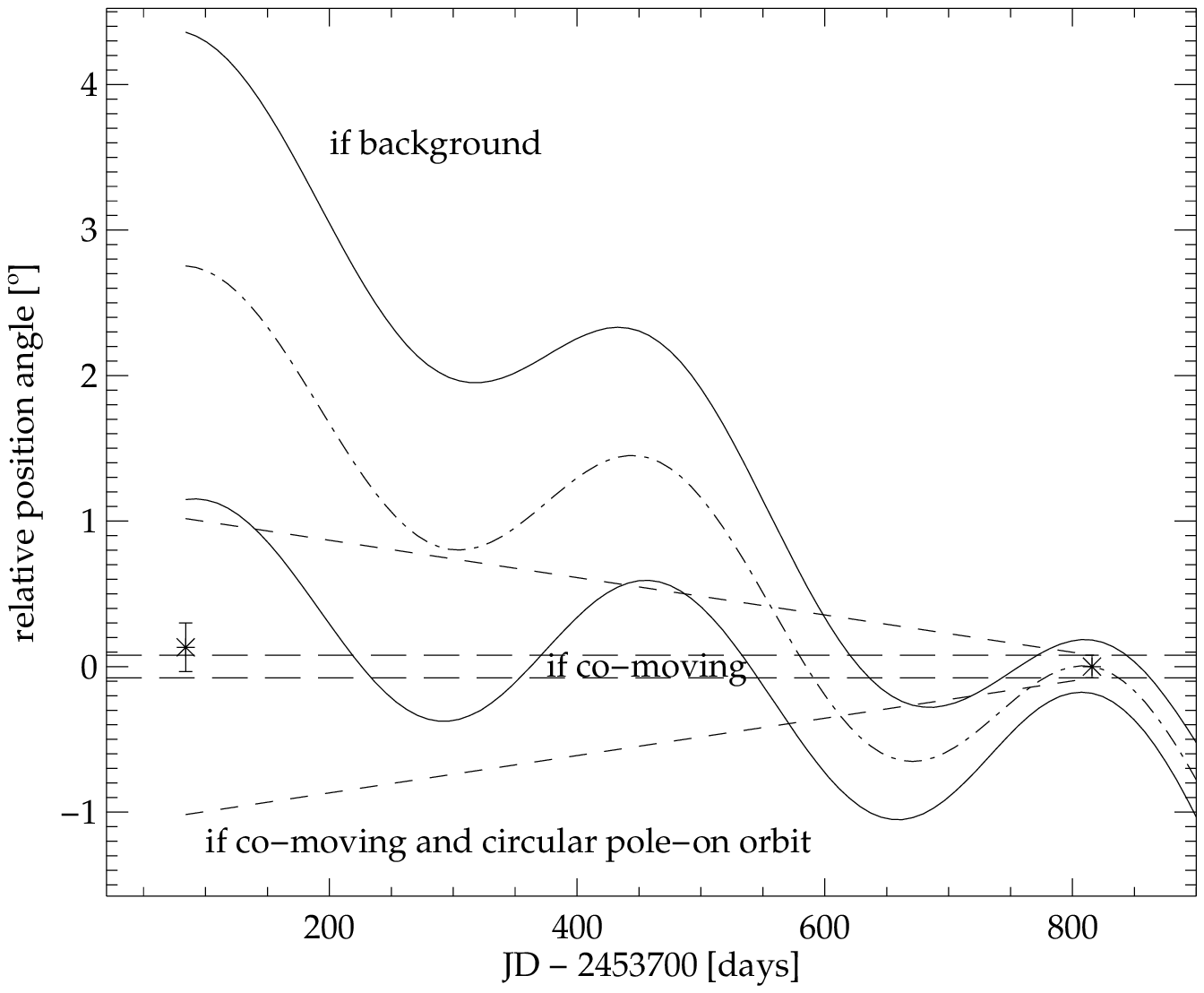}
   \caption{PMDs of Sz 22 from the relative astrometric measurements for separation (left) and position angle (right).}
   \label{FigSZ22PMD}
\end{figure*}

This target has hitherto been considered (spectroscopically) as a single star \citep{2003A&A...410..269M,2004AJ....127.1747H}, but it has a  known
companion candidate at an angular separation of $\sim$17.6 arcsec \citep{1997ApJ...481..378G}, outside of our field of view (FoV). Taking its strong interstellar
extinction of  A$_{\mathrm{v}}$\,$\sim$\,7 mag into account, its spectral type
corresponds to K6  \citep{2009ApJ...703.1964F,2007ApJS..173..104L}. This extinction and the Paschen $\beta$ emission \citep{2003AJ....125.2134G} could indicate the
presence of a
circumstellar disc, which is directly visible on images taken from the archive of the Hubble Space Telescope, as well as from our VLT NACO Ks band images (Fig. \ref{FigSZ22}).
This feature was described for the first time as a nebulous object with dimensions of 4'' x 7'' with the longer dimension at a position angle of
160$\degr$ in
\citet{1973ApJ...180..115H}. In the optical HST image, we find dimensions of 3'' x 5'' with the longer dimension at a position angle of $\sim$155$\degr$, while there seems
to be a slight change in orientation in the near-IR using NACO, where we find dimensions of 2.5'' x 5'' with the longer dimesion at a position angle of $\sim$145$\degr$.
The dimensions seem to shrink with time, while the orientation changes clockwise. However, \citet{2007A&A...462..211L} find out that the disc has a
relatively low mass ($\leq$ 0.009 M$_{\sun}$).

Nevertheless, this object could also be an embedded YSO surrounded by a compact reflection nebula having a large infrared excess, which shows a compact north-south flow HH 920 emerging
from it \citep{2006AJ....132.1923B}, consistent with its strong interstellar extinction of A$_{\mathrm{v}}$\,$\sim$7 mag despite its flat appearance, as well as the estimated age of
the star of $\leq$ 0.1 Myr \citep{2003AJ....125.2134G}. However, we find a best fitting age of about 1.4 Myr, based on
BCAH evolutionary models \citep{1998A&A...337..403B,2002A&A...382..563B}. For this purpose, the strong
interstellar extinction of A$_{\mathrm{v}}$\,$\sim$7 mag was transformed to K band according to the
interstellar extinction law values by \citet{1985ApJ...288..618R}, and the spectral type K6
\citep{2009ApJ...703.1964F,2007ApJS..173..104L} was converted to temperature based on stellar colours of Table A5 in
\citet{1995ApJS..101..117K}. In any case, Sz 22 is slightly below the average age of Cha I members, which is close to 2 Myr
\citep{2000A&A...359..269C}, or even one of the youngest members in this star-forming region.

The new stellar companion is located $\sim$0.5 arcsec west of the primary star, just outside the detected disc/reflection nebula (Fig. \ref{FigSZ22}).
Our PMDs (Fig. \ref{FigSZ22PMD}) confirm that both components are co-moving and that the companion increases its angular separation from
the primary by 2.5 mas each year, without changing
its position angle. Because the change in separation is close to the expectation for an edge-on orbit, we can conclude that the orbit is almost
edge-on, or else that the eccentricity is high. As the extended emission around the primary could either be a disc or a compact
reflection nebula, we cannot judge whether the disc around Sz 22 A and the probable edge-on orbit of Sz 22 A \& B are aligned.

The disc/reflection nebula and its strong extinction make it difficult to classify the spectral type of the stellar companion; its
J-Ks colours indicate an early M type, and its Ks band brightness is consistent with a spectral type of M4.5 $\pm$ 1.5 using the procedure described for the companion of DI Cha, if the
companion of Sz 22 is not extincted.

The wide visual companion candidate at $\sim$17.6 arcsec \citep{1997ApJ...481..378G} cannot be astrometrically proven by us, since it is outside of our FoV, but
it was shown to be binary by \citet{2008ApJ...683..844L}. According to the brightness ratio given there, we find spectral types of M2 $\pm$ 2 and M3 $\pm$ 2 using the
method described in the previous section. 

\onltab{5}{
\begin{table*}
\caption{Absolute astrometric results}
\label{table:5}
\begin{center}
\begin{tabular}{llcr@{\,$\pm$\,}lccr@{\,$\pm$\,}lccc}
\hline\hline
Object        & JD-2448000        & Ref. & \multicolumn{2}{c}{Separation} & Sign.$^a$  & Sign. orb. & \multicolumn{2}{c}{PA$^b$ }   & Sign.$^a$  & Sign. orb.\\
              & $[\mathrm{days}]$ &      & \multicolumn{2}{c}{[arcsec]}   & not Backg. & motion     & \multicolumn{2}{c}{[$\degr$]} & not Backg. & motion    \\
              &         & & $\rho$ & $\delta_{\rho}$ & $\sigma_{\rho,\,\mathrm{back}}$ & $\sigma_{\rho,\,\mathrm{orb}}$  
                        & $PA$   & $\delta_{PA}$   & $\sigma_{PA,\,\mathrm{back}}$   & $\sigma_{PA,\,\mathrm{orb}}$   \\
\hline
DI Cha AB                 & 5783.85826         &     & 4.5444  & 0.0609  & 0.5  & 0.1  & 202.375 & 1.236 & 0.4  & 0.1    \\
                          & 5821.80131         & 1   & 4.557   & 0.017   & 0.4  & 0.0  & 202.1   & 0.6   & 0.6  & 0.3    \\
                          & 6515.62668         &     & 4.5510  & 0.0689  & 0.2  & 0.0  & 202.542 & 1.400 & 0.1  & 0.0    \\
                          & 6882.64078         &     & 4.5556  & 0.0729  & $^c$ & $^c$ & 202.559 & 1.482 & $^c$ & $^c$   \\
\ \ \ \ \ \ \ \ 
\ \ \ \, AC               & 5783.85826         &     & 4.6015  & 0.0616  & 0.3  & 0.0  & 202.062 & 1.236 & 0.3  & 0.0    \\
                          & 6515.62668         &     & 4.6007  & 0.0696  & 0.1  & 0.0  & 202.136 & 1.400 & 0.1  & 0.0    \\
                          & 6882.64078         &     & 4.5966  & 0.0735  & $^c$ & $^c$ & 202.149 & 1.482 & $^c$ & $^c$   \\
\ \ \ \ \ \ \ \ 
\ \ \ \, A(BC)$^d$        & 1476.50000         & 2   & 4.9     & 0.2     & 0.7  & 1.5  & 202     & 3     & 0.8  & 0.3    \\
                          & 3578.23716         & 3,4 & 4.5926  & 0.0171  & 1.1  & 0.2  & 201.981 & 0.096 & 1.2  & 0.2    \\
                          & 5783.85826         &     & 4.5740  & 0.0613  & 0.4  & 0.0  & 202.206 & 1.236 & 0.3  & 0.1    \\
                          & 6515.62668         &     & 4.5762  & 0.0692  & 0.1  & 0.0  & 202.331 & 1.400 & 0.1  & 0.0    \\
                          & 6882.64078         &     & 4.5765  & 0.0732  & $^c$ & $^c$ & 202.345 & 1.482 & $^c$ & $^c$   \\
\ \ \ \ \ \ \ \ 
\ \ \ \, BC               & 5783.85826         &     & 0.0623  & 0.0012  & 1.8  & 4.5  & 178.613 & 1.661 & 6.2  & 6.4    \\
                          & 5821.80131         & 1   & 0.066   & 0.005   & 0.8  & 2.6  & 177.9   & 2.3   & 5.8  & 5.0    \\
                          & 6515.62668         &     & 0.0593  & 0.0018  & 1.0  & 2.5  & 169.131 & 1.579 & 4.5  & 2.3    \\
                          & 6882.64078         &     & 0.0524  & 0.0017  & $^c$ & $^c$ & 163.808 & 1.629 & $^c$ & $^c$   \\
\ \ \ \ \ \ \ \ 
\ \ \ \, AD               & 5783.85826         &     & 0.2071  & 0.0038  & 6.6  & 0.7  & 227.326 & 1.435 & 4.2  & 3.1    \\
                          & 6515.62668         &     & 0.2100  & 0.0046  & 2.6  & 0.2  & 231.638 & 1.620 & 1.6  & 1.0    \\
                          & 6882.64078         &     & 0.2105  & 0.0038  & $^c$ & $^c$ & 233.890 & 1.529 & $^c$ & $^c$   \\
Sz 22 AB                  & 5783.88427         &     & 0.5107  & 0.0068  & 3.8  & 0.5  & 271.709 & 1.236 & 1.1  & 0.1    \\
                          & 6515.64511         &     & 0.5157  & 0.0078  & $^c$ & $^c$ & 271.577 & 1.402 & $^c$ & $^c$   \\
CHXR 32 AB                & 5819.91198         & 6   & 2.4372  & 0.0330  & 0.9  & 0.2  & 285.011 & 1.244 & 0.2  & 0.0    \\
                          & 6882.90637         &     & 2.4471  & 0.0391  & $^c$ & $^c$ & 284.924 & 1.482 & $^c$ & $^c$   \\
\ \ \ \ \ \ \ \ 
\ \ \ \ \ \ \ \ 
\ AC                      & 5819.91198         & 6   & 2.3681  & 0.0345  & 1.4  & 0.3  & 285.311 & 1.249 & 0.0  & 0.2    \\
                          & 6882.90637         &     & 2.3534  & 0.0377  & $^c$ & $^c$ & 285.649 & 1.482 & $^c$ & $^c$   \\
\ \ \ \ \ \ \ \ 
\ \ \ \ \ \ \ \ 
\ A(BC)$^d$               & 1476.50000         & 2   & 2.5     & 0.5     & 0.7  & 0.1  & 284     & 5     & 0.1  & 0.2    \\
                          & 4690.70849         & 5   & 2.430   & 0.002   & 2.8  & 0.1  & 285.1   & 0.5   & 0.4  & 0.0    \\
                          & 5819.91198         & 6   & 2.4215  & 0.0329  & 1.1  & 0.1  & 285.077 & 1.244 & 0.2  & 0.0    \\
                          & 6516.89621         &     & 2.4292  & 0.0369  & 0.5  & 0.1  & 285.103 & 1.401 & 0.1  & 0.0    \\
                          & 6882.90637         &     & 2.4257  & 0.0388  & $^c$ & $^c$ & 285.084 & 1.482 & $^c$ & $^c$   \\
\ \ \ \ \ \ \ \ 
\ \ \ \ \ \ \ \ 
\ A(BC)$^e$               & 5819.91198         & 6   & 2.4095  & 0.0329  & 1.1  & 0.0  & 285.129 & 1.245 & 0.1  & 0.0    \\
                          & 6882.90637         &     & 2.4096  & 0.0385  & $^c$ & $^c$ & 285.213 & 1.482 & $^c$ & $^c$   \\
\ \ \ \ \ \ \ \ 
\ \ \ \ \ \ \ \ 
\ BC                      & 5819.91198         & 6   & 0.0702  & 0.0109  & 0.7  & 2.5  & 94.834  & 2.801 & 3.1  & 2.4    \\
                          & 6882.90637         &     & 0.0984  & 0.0016  & $^c$ & $^c$ & 87.322  & 1.497 & $^c$ & $^c$   \\
Cha H$\alpha$ 5 AB        & 5782.75127         &     & 0.0481  & 0.0023  & 1.4  & 0.6  & \ 88.244& 4.837 & 8.4  & 2.6    \\
                          & 6881.73805         &     & 0.0458  & 0.0028  & $^c$ & $^c$ & 111.840 & 7.758 & $^c$ & $^c$   \\
\ \ \ \ \ \ \ \ 
\ \ \ \ \ \ \ \, Acc1     & 5782.75127         &     & 1.4104  & 0.0191  & 0.1  & 0.8  & 222.108 & 1.243 & 0.9  & 0.7    \\
                          & 6881.73805         &     & 1.3884  & 0.0223  & $^c$ & $^c$ & 220.738 & 1.483 & $^c$ & $^c$   \\
\ \ \ \ \ \ \ \ 
\ \ \ \ \ \ \ \, Bcc1     & 5782.75127         &     & 1.4442  & 0.0195  & 0.3  & 1.3  & 223.484 & 1.240 & 1.0  & 0.5    \\
                          & 6881.73805         &     & 1.4040  & 0.0228  & $^c$ & $^c$ & 222.508 & 1.490 & $^c$ & $^c$   \\
\ \ \ \ \ \ \ \ 
\ \ \ \ \ \ \ \, 
(AB)cc1$^f$               & 5782.75127       &     & 1.4272  & 0.0194  & 0.1  & 1.0  & 222.804 & 1.244 & 0.9  & 0.6    \\
                          & 6516.68649         &     & 1.4103  & 0.0213  & 0.1  & 0.4  & 222.077 & 1.400 & 0.3  & 0.2    \\
                          & 6881.73805         &     & 1.3963  & 0.0225  & $^c$ & $^c$ & 221.619 & 1.486 & $^c$ & $^c$   \\
\hline
\end{tabular}
\end{center}
\textbf{Notes.} $^{(a)}$ Assuming the fainter component is a non-moving background star. $^{(b)}$ Position Angle (PA) is measured
from N over E to S. $^{(c)}$ Significances are given relative to the last epoch. $^{(d)}$ Results of component A relative to the 
centre of brightness of components B and C. $^{(e)}$ Results of component A relative to the centre of mass (masses from apparent
magnitudes, Table \ref{table:9}, and distance of Chamaeleon cloud of 165 $\pm$ 30 pc, using the models of 
\citet{1998A&A...337..403B}, giving for B  0.76 M$_{\sun}$ and for C 0.28 M$_{\sun}$ at 2\,Myr) of components B and C.
$^{(f)}$ Results of the centre of brightness of components A and B relative to the further companion candidate cc1.\\
\textbf{References.} (1) \citet{2008ApJ...683..844L}. (2) \citet{1997ApJ...481..378G}; date given, assuming midnight for
calculation of JD. (3) HST data from ESO/ST-ECF science archive, only position measurement error in RA and Dec considered. 
(4) From ESO/ST-ECF science archive, see also \citet{2001ASPC..231..620S}, \citet{2001AAS...199.6009P}.
(5) \citet{2006A&A...459..909C}. (6) Rereduced, see also \citet{2008ApJ...683..844L}.
\end{table*}
}

\onltab{6}{
\begin{table*}
\caption{Relative astrometric results}
\label{table:6}
\begin{center}
\begin{tabular}{llr@{\,$\pm$\,}lccr@{\,$\pm$\,}lcc}
\hline\hline
Object & Epoch differ- & \multicolumn{2}{c}{Change in sepa-} & Sign.$^{a,c}$ & Sign.$^c$ orb. & \multicolumn{2}{c}{Change in}        & Sign.$^{a,c}$  & Sign.$^c$ orb.\\
       & ence [days]   & \multicolumn{2}{c}{ration [pixel]}  & not Backg.    & motion         & \multicolumn{2}{c}{PA$^b$ [$\degr$]} & not Backg.     & motion        \\
       & \ \ \ \ \ \ \ \ $\Delta$t     
            & $\Delta\,\rho$ & $\delta_{\Delta\,\rho}$ & $\sigma_{\rho,\,\mathrm{back}}$ & $\sigma_{\rho,\,\mathrm{orb}}$  
            & $\Delta\,PA$ & $\delta_{\Delta\,PA}$ & $\sigma_{PA,\,\mathrm{back}}$ & $\sigma_{PA,\,\mathrm{orb}}$   \\
\hline
DI Cha AB                 & 1098.78252     &  0.846    & 0.928 & 3.3     & 0.9  &  0.184    & 0.249 & 2.6     & 0.7 \\
                          & \ \ 367.01410  &  0.347    & 0.379 & 2.8     & 0.9  &  0.017    & 0.092 & 1.9     & 0.2 \\
\ \ \ \ \ \ \ \ 
\ \ \ \, AC               & 1098.78252     & -0.365    & 0.942 & 2.0     & 0.4  &  0.087    & 0.249 & 2.3     & 0.3 \\
                          & \ \ 367.01410  & -0.302    & 0.373 & 1.3     & 0.8  &  0.013    & 0.091 & 1.9     & 0.1 \\
\ \ \ \ \ \ \ \ 
\ \ \ \, A(BC)$^d$        & 1098.78252     &  0.187    & 0.942 & 2.6     & 0.2  &  0.139    & 0.249 & 2.5     & 0.6 \\
                          & \ \ 367.01410 & 0.020 & 0.389 & 2.0 & 0.1 &  0.014    & 0.092 & 1.9     & 0.2 \\
\ \ \ \ \ \ \ \ 
\ \ \ \, BC               & 1098.78252     & -0.743    & 0.131 & 2.0     & 5.7  & -14.805   & 1.325 & 6.9     & 11  \\
                          & \ \ 367.01410  & -0.518    & 0.163 & 1.1     & 3.2  & -5.323    & 1.001 & 6.6     & 5.3 \\
\ \ \ \ \ \ \ \ 
\ \ \ \, AD               & 1098.78252     &  0.256    & 0.238 & 8.5     & 1.1  &  6.564    & 0.859 & 7.0     & 7.6 \\
Sz 22 AB                  & \ \ 731.76084  &  0.374    & 0.085 & 5.2     & 4.4  & -0.132    & 0.184 & 1.6     & 0.7 \\
CHXR 32 AB                & 1062.99439     &  0.746    & 0.529 & 3.3     & 1.4  & -0.087    & 0.243 & 1.0     & 0.4 \\
\ \ \ \ \ \ \ \ 
\ \ \ \ \ \ \ \ 
\ AC                      & 1062.99439     & -1.109    & 1.092 & 3.8     & 1.0  &  0.338    & 0.268 & 0.1     & 1.3 \\
\ \ \ \ \ \ \ \ 
\ \ \ \ \ \ \ \ 
\ A(BC)$^d$               & 1062.99439     &  0.322    & 0.557 & 3.6     & 0.6  &  0.007    & 0.244 & 0.8     & 0.0 \\
                          & \ \ 366.01016  & -0.260    & 0.336 & 3.8     & 0.8  & -0.019    & 0.103 & 0.8     & 0.2 \\
\ \ \ \ \ \ \ \ 
\ \ \ \ \ \ \ \ 
\ A(BC)$^e$               & 1062.99439     & 0.006     & 0.613 & 3.8     & 0.0  &  0.084    & 0.246 & 0.5     & 0.3 \\
\ \ \ \ \ \ \ \ 
\ \ \ \ \ \ \ \ 
\ BC                      & 1062.99439     &  2.130    & 0.820 & 0.7     & 2.6  & -7.512    & 2.530 & 3.2     & 3.0 \\
Cha H$\alpha$ 5 AB        & 1098.98678     & -0.170    & 0.259 & 1.4     & 0.7  & 23.596    & 8.940 & 8.5     & 2.6 \\
\ \ \ \ \ \ \ \ 
\ \ \ \ \ \ \ \, Acc1     & 1098.98678     & -1.662    & 0.395 & 0.2     & 4.2  & -1.370    & 0.295 & 1.8     & 4.6 \\
\ \ \ \ \ \ \ \ 
\ \ \ \ \ \ \ \, Bcc1     & 1098.98678     & -3.037    & 0.476 & 0.4     & 6.4  & -0.976    & 0.315 & 2.1     & 3.1 \\
\ \ \ \ \ \ \ \ \ \ \
\ \ \ \ \, (AB)cc1$^f$    & 1098.98678     & -2.333    & 0.456 & 0.1     & 5.1  & -1.185    & 0.313 & 1.9     & 3.8 \\
                          & \ \ 365.05156 & -1.055 & 0.257 & 0.5 & 4.1  & -0.458    & 0.153 & 1.7     & 3.0 \\
\hline                                                   
\end{tabular}
\end{center}
\textbf{Notes.} $^{(a)}$ Assuming the fainter component is a non-moving background star. $^{(b)}$ PA is measured from N over E to S. $^{(c)}$ Significances are given relative to the last epoch.
$^{(d)}$ Results of component A relative to the centre of brightness of components B and C. $^{(e)}$ See $^{e}$ in Table \ref{table:5}.
$^{(f)}$ Results of the centre of brightness of components A and B relative to the further companion candidate cc1.
\end{table*}
}

\subsection{The triple system CHXR 32}

\begin{figure*}
\centering
\resizebox{0.65\textwidth}{!}{\includegraphics{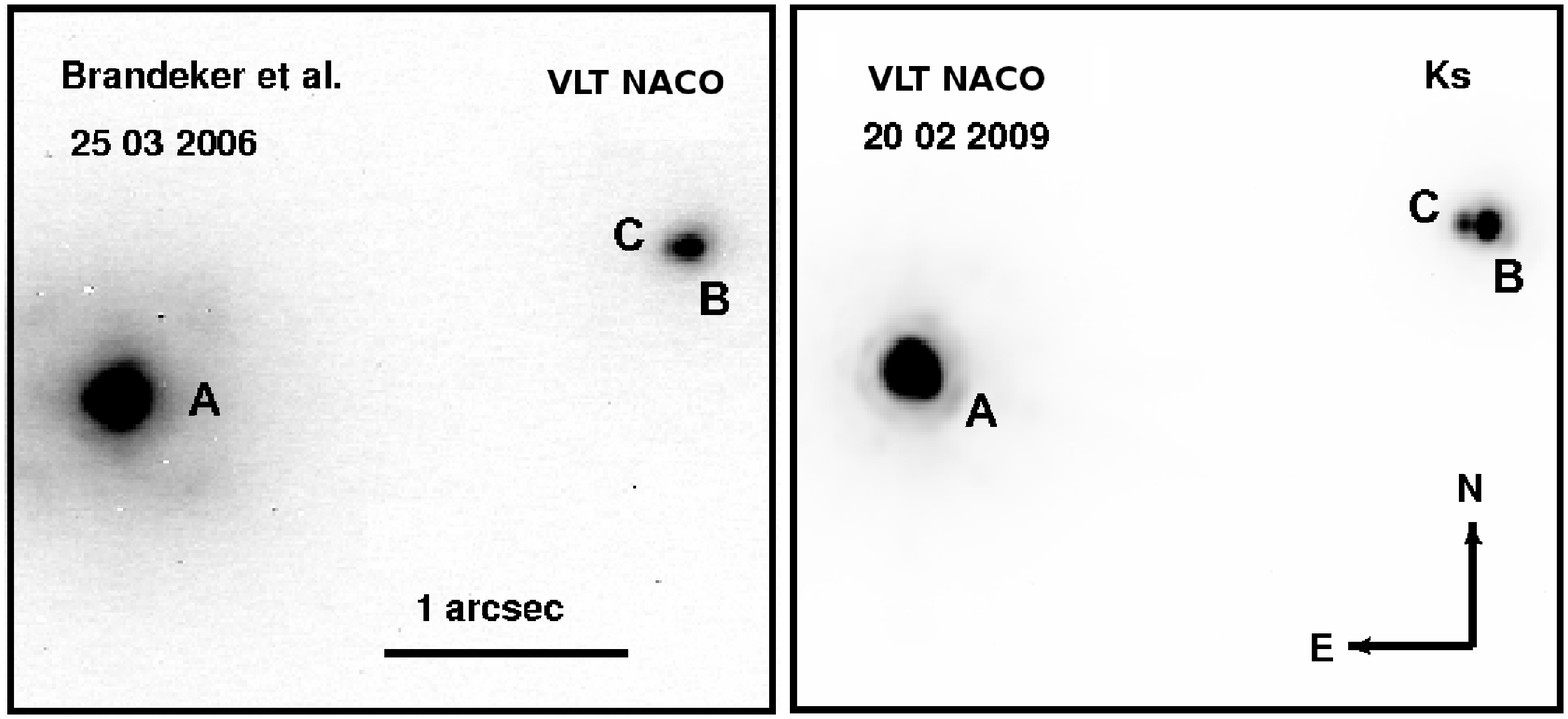}}
\caption{VLT NACO Ks band images of CHXR 32. The left frame shows the first observation by Brandeker et al. \citep{2008ApJ...683..844L} as reduced
again by us with an elongated PSF of the
component B/C. The right frame, taken by us nearly three years later as a second epoch, confirms the triple nature of this target, showing two resolved PSFs of B and C.}
\label{FigCHXR32}
\end{figure*}

\begin{figure*}
   \centering
   \includegraphics[width=0.39\textwidth]{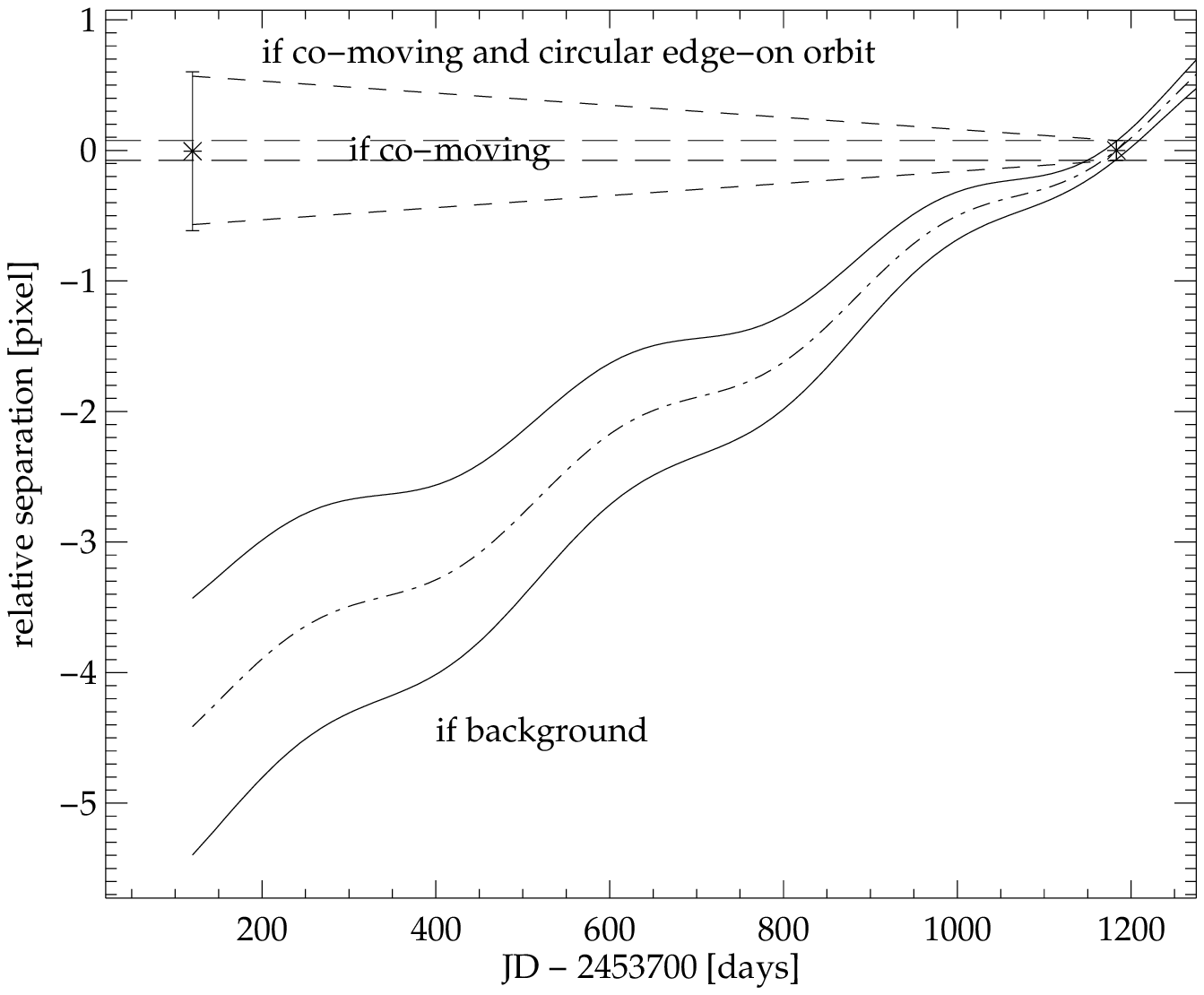}
   \includegraphics[width=0.39\textwidth]{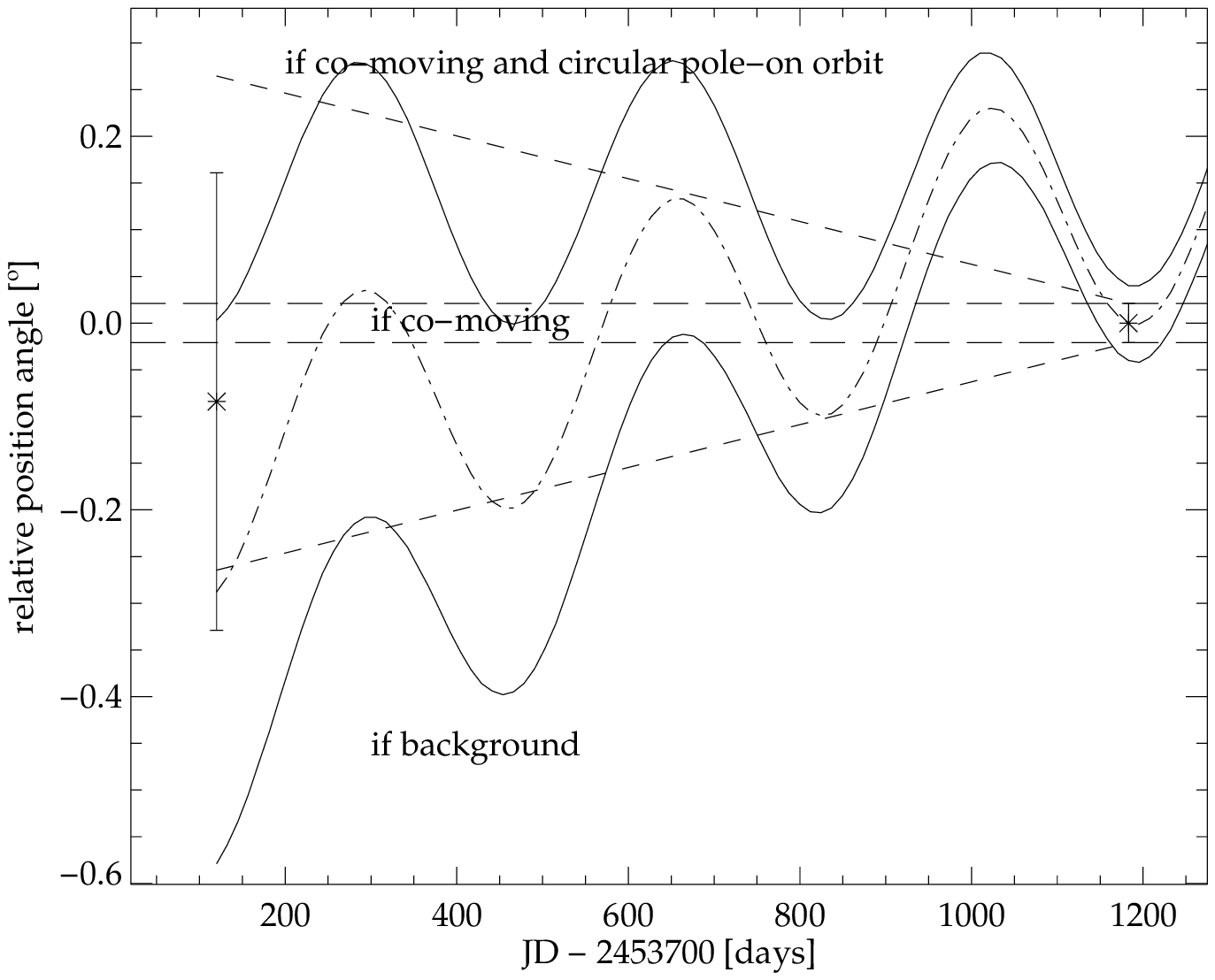}
   \includegraphics[width=0.39\textwidth]{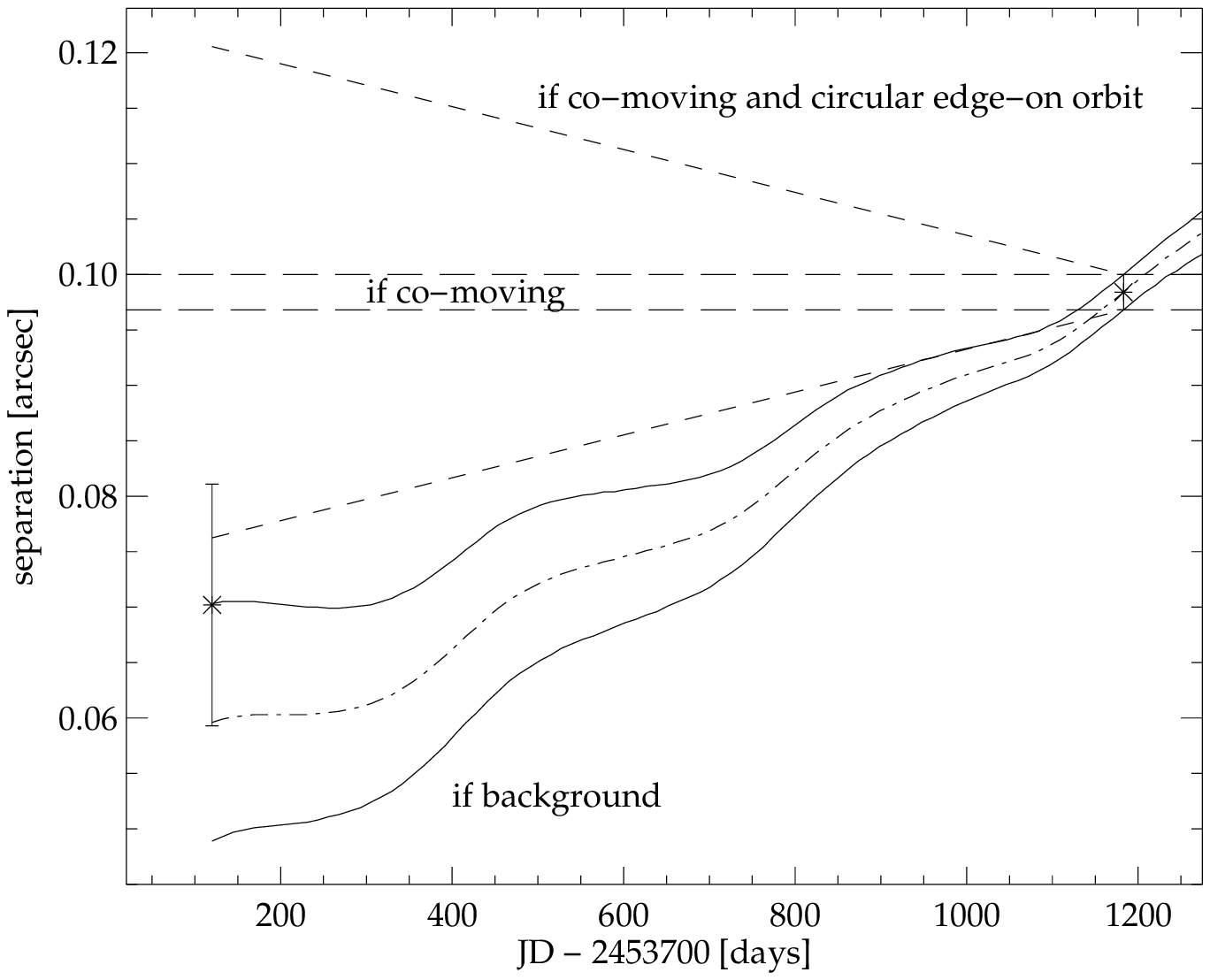}
   \includegraphics[width=0.39\textwidth]{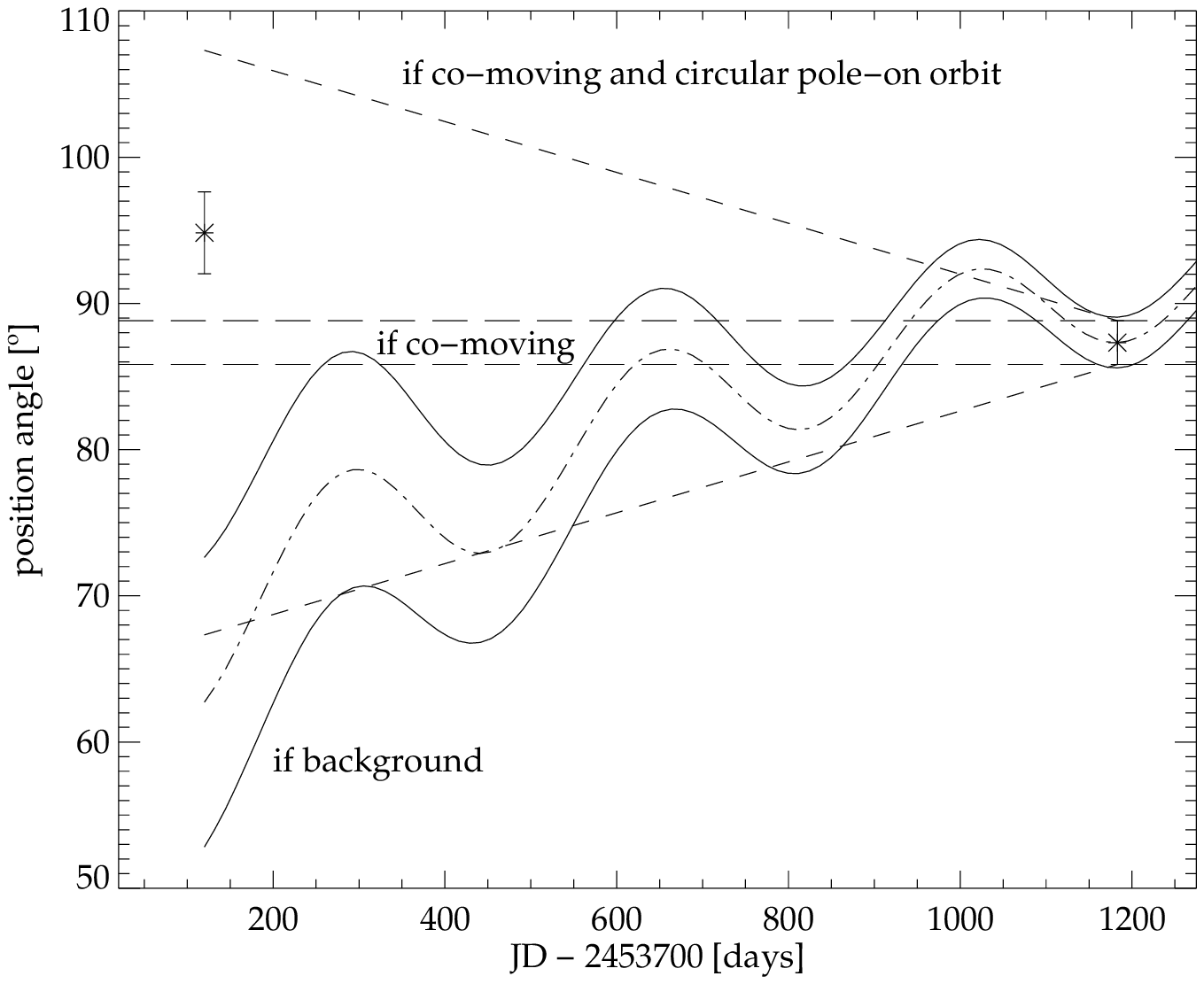}
   \caption{PMDs of CHXR 32 from the relative and absolute astrometric measurements for separation (left) and position angle (right).
The upper panels refer to the centre of mass of the pair B/C (adopting that the mass of B is 1.5 times that of C), relative to the main component A. The separation PMD rejects here the
background hypothesis. The lower panels are PMDs of C relative to B, assuming that the proper motion of B is identical to that of A.  Here the background hypothesis
is excluded by the position angle PMD.}
   \label{FigCHXR32PMD}
\end{figure*}

According to \citet{2003A&A...410..269M} and \citet{2012ApJ...745..119N}, CHXR 32 is spectroscopically single. 
However, this star, also named Glass I, is a well known binary of about 2.4 arcsec separation \citep{1988A&A...207...46C}. 
We here follow the notation of \citet{2006A&A...459..909C}, naming as ``A'' the
eastern component, although it is fainter than ``B'' at optical wavelengths owing to a strong circumstellar extinction. CHXR 32 A is a G-type emission
line star (H$\alpha$ emission equivalent width $<$\,10 \AA, hence a WTTS), possibly surrounded by a disc that diminishes the optical flux. For component B a spectral
type K4 was determined \citep[][there denominated as ``A'']{1993A&A...278...81R}. 

When we again reduced the data used by us as our first epoch \citep[but originally from][]{2008ApJ...683..844L}, we noticed that the object
B had an elongated shape. Using IDL Starfinder \citep{2000SPIE.4007..879D}, we fitted the point-spread-function (PSF) of A to this component, revealing the best fit for a double
source at this position. To confirm this and to search for common proper motion, we obtained a new NACO image in 2009 that clearly separates
the two components of B, with only 0.08 arcsec angular separation, now called B and C (Fig. \ref{FigCHXR32}). The
corresponding PMDs (Fig. \ref{FigCHXR32PMD}) confirm that all three components are co-moving. There are no significant orbital movements in the wide pair A-BC; however,
there are strong indications of orbital variations in position angle and especially in the separation between B and C, which increases at a rate of nearly 10 mas/year,
facilitating the direct detection of this binary in 2009 (Tables \ref{table:7} \& \ref{table:8}). From the strong change in separation close to the expectation for an edge-on orbit, we can
conclude that the orbit is close to edge-on, or that the eccentricity is high.
Although the PMD of the separation variation between B and C is compatible with the background hypothesis, the variations in the position angle are not
compatible with those of a background star. 

As for the other targets, we used the brightnesses of the primaries to determine mean apparent magnitudes of the secondaries using the measured contrasts and assuming no significant
variations; however, as for CHXR 32 A, 2MASS only gives upper limits in the H and K bands, so we use the
brightnesses of H = 7.189 $\pm$ 0.114 mag and Ks = 6.182 $\pm$ 0.089 mag given in \citet{2002AJ....124.1001C}, reflecting
the strong extinction of the primary. According to the photometry (Table \ref{table:9}) the secondary component of this now resolved triple star consists of an M1 (K6\,--\,M3.5) type
star with an M3.5 (M2.5\,--\,M5.5) companion, compatible with the unresolved H-Ks colour measured for B \& C. We note, however, that there are strong systematic errors
because the J band brightness varies by 2.3 $\sigma$ between 2MASS and \citet{2002AJ....124.1001C}, while the visible light curve varies up to about 2 mag according to ASAS
data presented in \citet{2013ApJ...764..127K}.

Since the B\,\&\,C components could not be fitted individually using PSF fitting in the J band data from 2006 and Ks band data from 2008 (Table \ref{table:9}), we used the
dedicated aperture photometry routines from ESO-MIDAS \citep{1992ASPC...25..120B} to measure a combined brightness difference relative to component A, because PSF fitting
using the single PSF of CHXR 32 A as comparison leads to overestimated brightness differences owing to a slight elongation of the
combined light of B\,\&\,C.

For the creation of the PMDs mentioned above, we decided to use the proper motion of UCAC2 \citep{2004AJ....127.3043Z} instead of UCAC3 \citep{2010AJ....139.2184Z}, since
this proper motion value is consistent with the proper motion of Cha I. The present discrepancy in UCAC3 can most likely be explained by the inclusion of 140 catalogues of all
kinds in different
wavelength bands. If epochs from near-infrared catalogues, in which the extincted CHXR 32 A is dominant, are mixed with those from optical catalogues, in which
CHXR 32 BC are the dominant sources, the spatial separation of $\sim$2.4 arcsec between A \& BC can be misinterpreted as a deviation from the real proper motion within the epoch
differences of several years. 
The strong photometric variability in ASAS data between 2002 to 2004 in CHXR 32 A reported in
\citet{2013ApJ...764..127K}, could lead to further discrepancies.
Please see this publication for further discussion of the proper motion behaviour of the system.

\subsection{The unresolved binary Cha H$\alpha$ 5}

\begin{figure*}
\centering
\resizebox{0.7\textwidth}{!}{\includegraphics{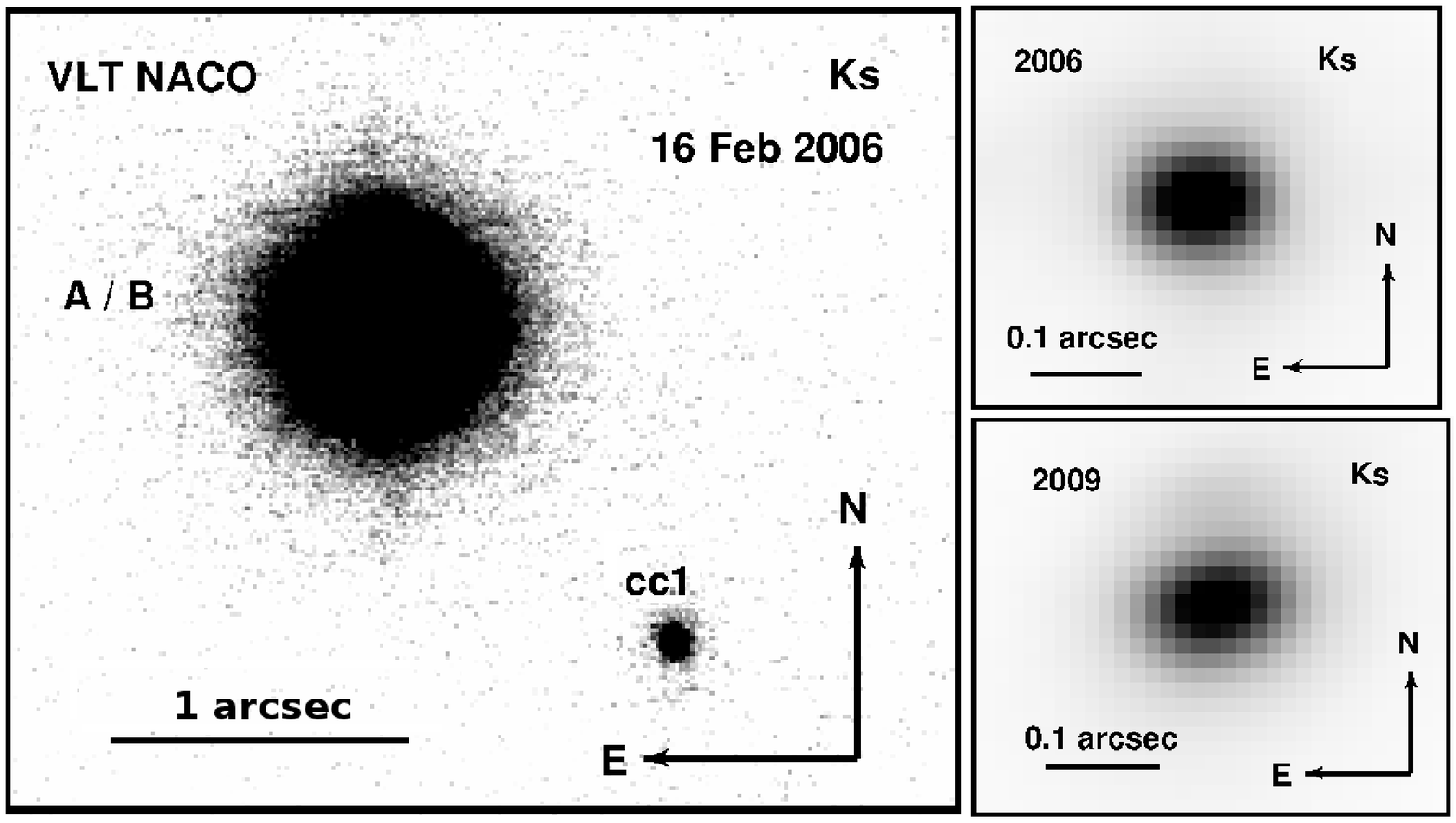}}
\caption{VLT NACO Ks band images of Cha H$\alpha$ 5. \textit{Left:} Our first observation in 2006 revealed a slightly elongated PSF of the primary (designated A/B), compared to the PSF of the
much fainter background object cc1. \textit{Right:} the position angle (orientation) of the elongation caused by the presence of two components has changed from the epochs in 2006 to 2009,
confirming orbital motion of this (still unresolved) binary.}
\label{FigChaHa5}
\end{figure*}

\begin{figure*}
   \centering
   \includegraphics[width=0.39\textwidth]{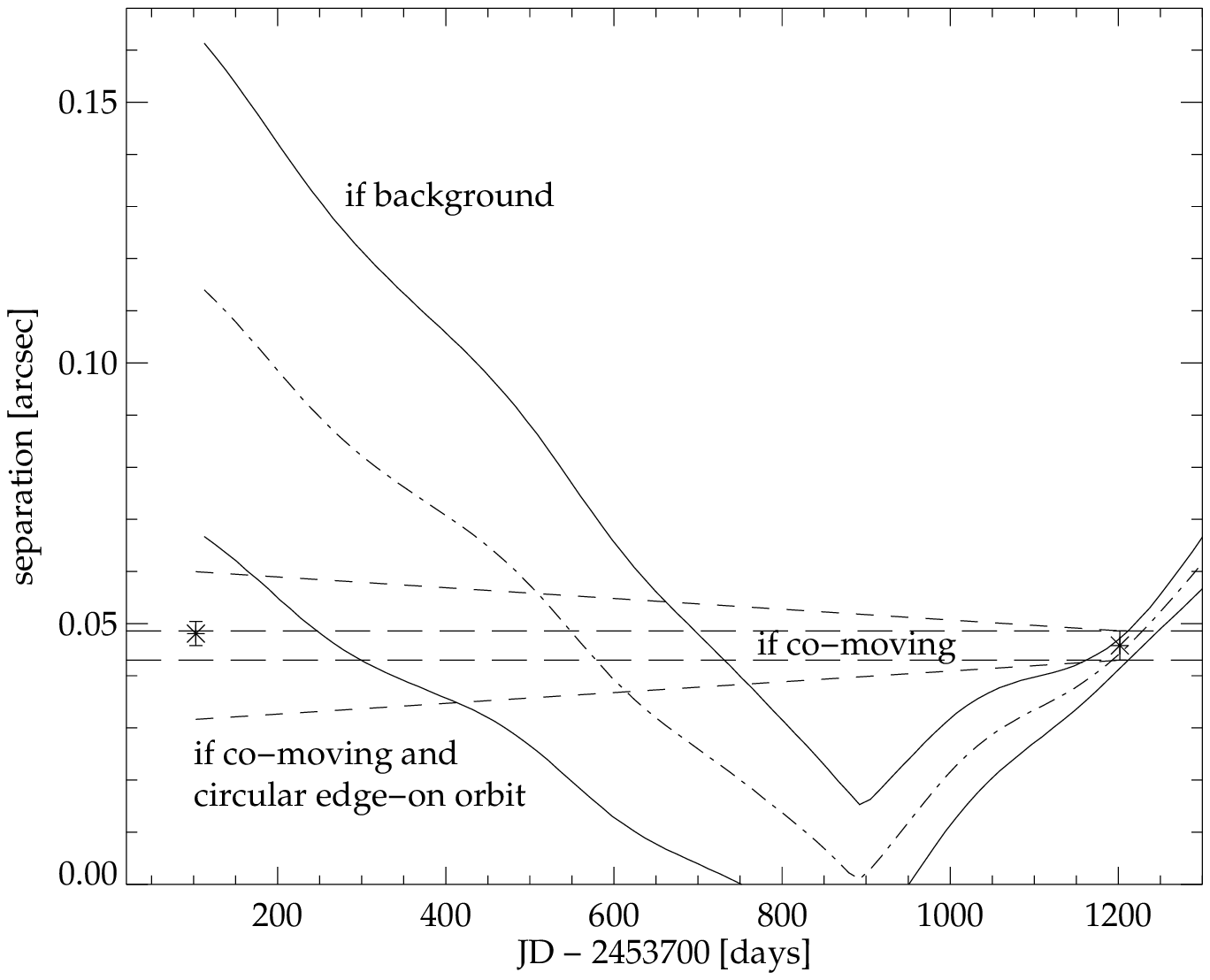}
   \includegraphics[width=0.39\textwidth]{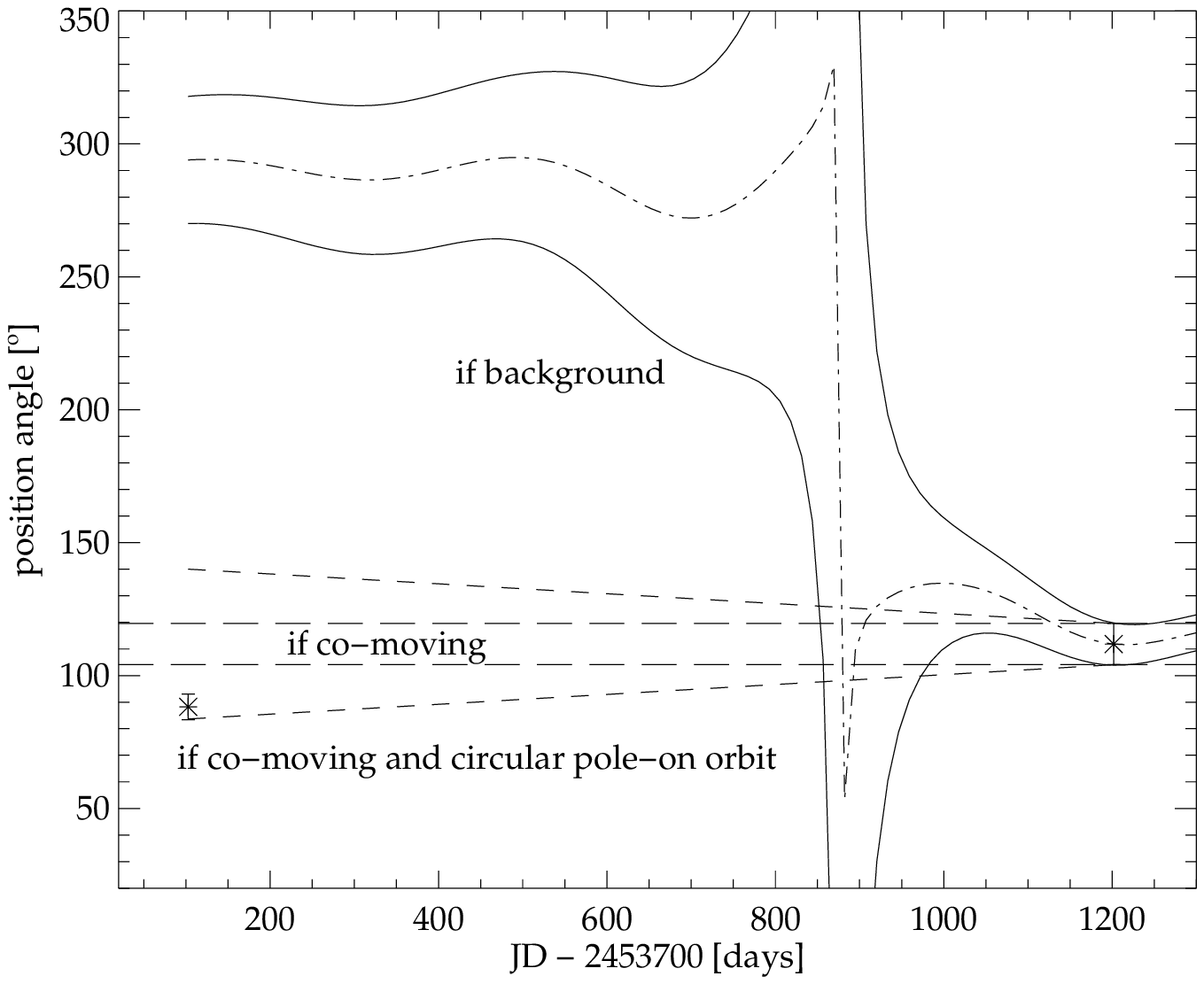}
   \includegraphics[width=0.39\textwidth]{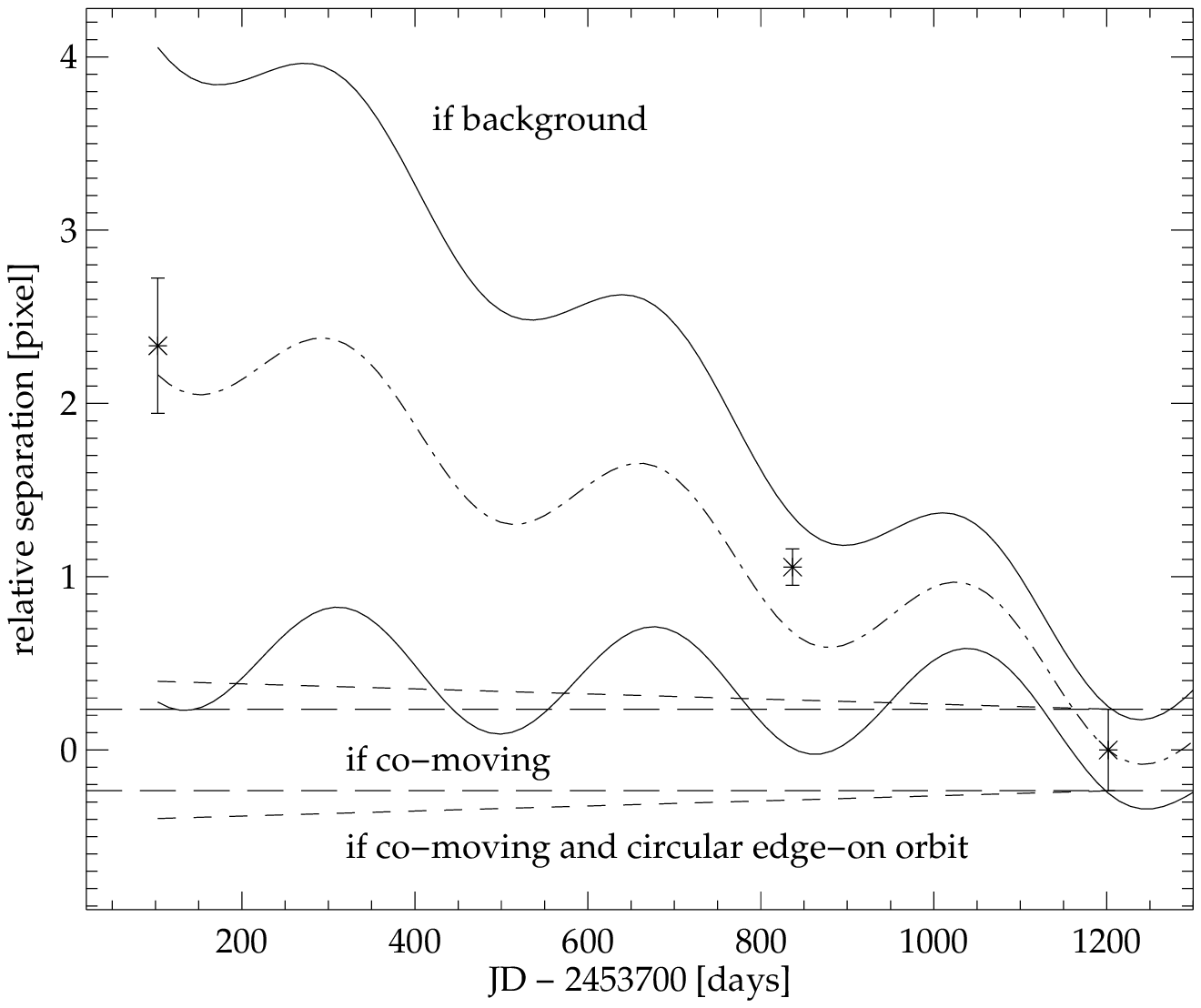}
   \includegraphics[width=0.39\textwidth]{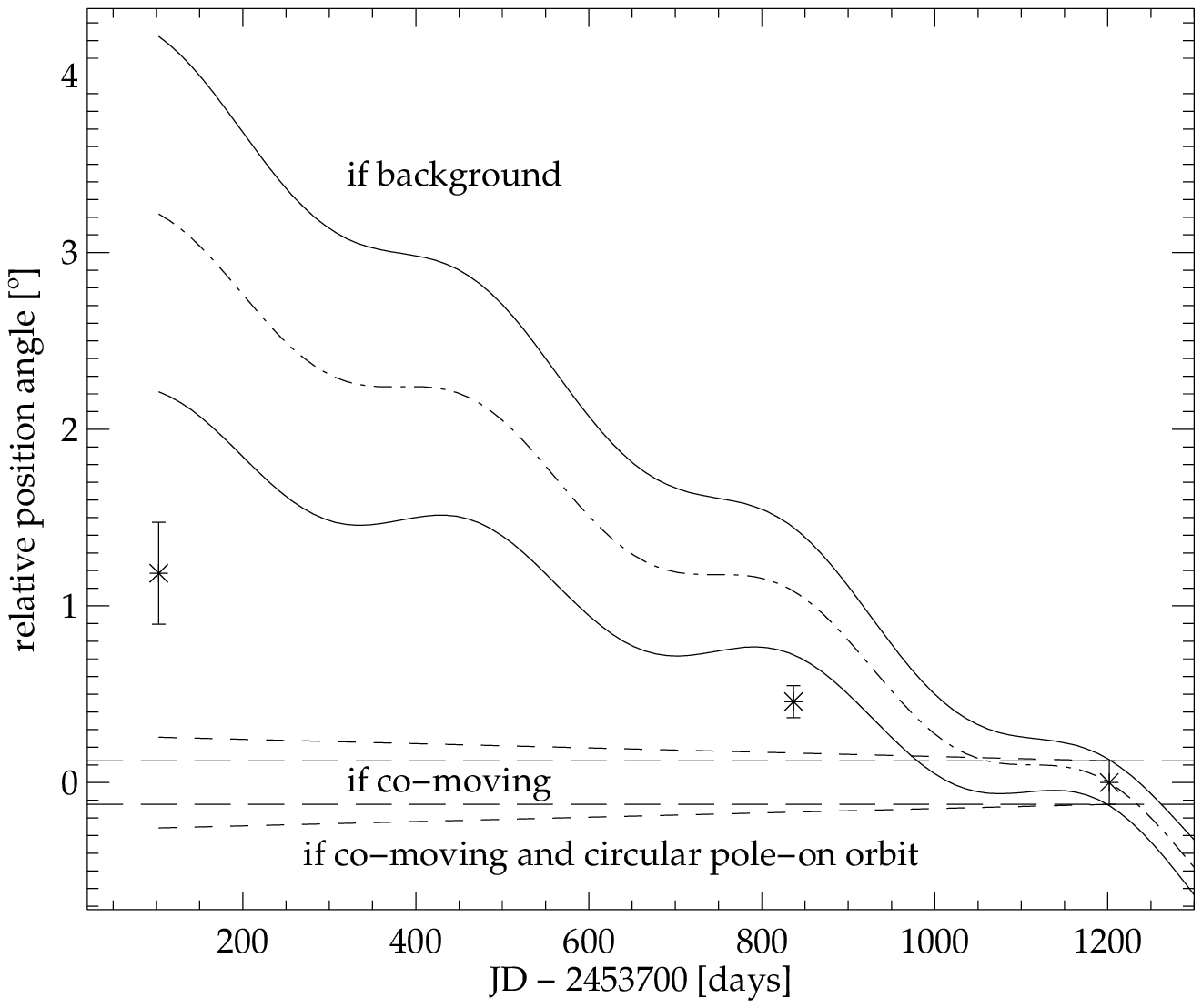}
   \caption{PMDs of Cha H$\alpha$ 5 from the absolute and relative astrometric measurements for separation (left) and position angle (right). The upper panels refer to
the co-moving A/B pair, the lower ones to the background object cc1 relative to A.}
   \label{FigChaHa5PMD}
\end{figure*}

This target attracted early attention after \citet{2002A&A...384..999N} found a candidate for a sub-stellar companion just 1.5'' away from
Cha H$\alpha$ 5. Based on its low apparent luminosity, this object could even have been of planetary mass. However, applying optical (FORS1) and infrared
(ISAAC) spectroscopy, \citet{2003IAUS..211..309N} show that it is a background object.
Here, we confirm this conclusion for the first time based on astrometric observations at three epochs (Figs. \ref{FigChaHa5} \& \ref{FigChaHa5PMD}). 

During this investigation we noticed that Cha H$\alpha$ 5 had elongated images (Fig. \ref{FigChaHa5}), while the PSF of the background object appeared perfectly
circular. This should not be the case if the elongation was caused by problems with the adaptive optics, especially because
Cha H$\alpha$ 5 was used as AO guide star. Therefore, we fitted the PSF of the background object to the images of Cha H$\alpha$ 5, obtained in 2006 and 2009
(those of 2008 had too bad a quality owing to adverse weather conditions), using the IDL Starfinder software \citep{2000SPIE.4007..879D}. In both cases the best
fit revealed two components of only $\sim$0.05 arcsec separation, just below the theoretical resolution limit of NACO/UT4 of 68 mas.

The PMDs (Fig. \ref{FigChaHa5PMD}) show that this close pair is co-moving. There is no significant change in the angular separation between 2006 and 2009, but a strong
variation in the position angle, corresponding to
an orbital period of about 46 years. This is consistent with two stellar components of 0.12 M$_{\sun}$ each, according to Kepler's third law. The projected separation between both
components is about 8 AU (adopting 165 pc distance from the Sun), just outside the range in which \citet{2008A&A...492..545J} could find companions with the radial velocity method,
in this way linking both methods and thus avoiding any “blind” gap of detection possibility.
The short period, visible only in the position angle (circular pole-on orbit), will soon enable confirmation of a curved orbital motion.

The resolved brightnesses of Cha H$\alpha$ 5 A \& B are compatible with spectral types of M6.5 (M5\,--\,M7) for both components, using the BCAH and COND evolutionary models, and are
thus fully consistent with M6 \citep{1998Sci...282...83N} and M6.5 \citep{2004A&A...416..555L} for an assumed age of 2 Myr, which most of the members
of Cha I are \citep{2000A&A...359..269C}, but in contrast to earlier estimates of 0.4 Myr \citep{2003AJ....125.2134G}. 
There seems to be some resemblance to the Cha H$\alpha$ 2 system \citep{2008A&A...484..413S}, but with a much shorter orbital period. The distances from the Sun of both systems are comparable,
while the magnitudes are only slightly fainter, therefore we cannot exclude that Cha H$\alpha$ 5 B and even A are brown dwarfs.
The corresponding 1\,$\sigma$ error minimum masses of the components are 68 \& 65 M$_{\mathrm{Jup}}$ according to COND models, respectively.

As given in the previous section for part of the photometry of CHXR 32 A(BC), we can only measure a brightness difference of cc1 with respect to the combined light of components A\,\&\,B in the data
of 2008, using aperture photometry with ESO-MIDAS. 

\begin{table}
\caption{Projected orbital separations and linear orbital movement fit results from absolute astrometric measurements }
\label{table:7}
\begin{center}
\begin{tabular}{lcr@{\,$\pm$\,}lr@{\,$\pm$\,}l}
\hline\hline
Object & proj.  & \multicolumn{2}{c}{Change in  } & \multicolumn{2}{c}{Change     }  \\
       & sep.   & \multicolumn{2}{c}{separation} & \multicolumn{2}{c}{in PA$^a$   }  \\
       & [AU]   & \multicolumn{2}{c}{[mas/yr]  } & \multicolumn{2}{c}{[$\degr$/yr]}  \\
\hline
DI Cha A(BC)$^b$          & 755       & -3.70      & 5.45 &  0.04 & 0.10 \\
\ \ \ \ \ \ \ \ 
\ \ \ \, BC               & \ \ 10    & -2.99      & 0.65 & -4.86 & 0.68 \\
\ \ \ \ \ \ \ \ 
\ \ \ \, AD               & \ \ 35    &  1.16      & 1.76 &  2.18 & 0.68 \\
Sz 22 AB                  & \ \ 85    &  2.50      & 5.17 & -0.07 & 0.93 \\
CHXR 32 A(BC)$^b$         & 401       & -0.91      & 4.43 &  0.01 & 0.18 \\
\ \ \ \ \ \ \ \ 
\ \ \ \ \ \ \ \ 
\ A(BC)$^c$               & 398       &  0.03      & 17.4 &  0.03 & 0.67 \\
\ \ \ \ \ \ \ \ 
\ \ \ \ \ \ \ \ 
\ BC                      & \ \ 14    &  9.69      & 3.79 & -2.58 & 1.09 \\
Cha H$\alpha$ 5 AB        & \ \ \ \ 8 & -0.76      & 1.20 &  7.84 & 3.04 \\
\ \ \ \ \ \ \ \ \ \ \
\ \ \ \ \, (AB)cc1$^d$    & (232)     & -9.90      & 9.51 & -0.38 & 0.62 \\
\hline
\end{tabular}
\end{center}
\textbf{Notes.} $^{(a)}$ PA is measured from N over E to S. 
$^{(b)}$ See $^{d}$ in Table \ref{table:5}. $^{(c)}$ See $^{e}$ in Table \ref{table:5}. $^{(d)}$ See $^{f}$ in Table \ref{table:5}.
\end{table}

\begin{table}
\caption{Projected orbital separations and linear orbital movement fit results from relative astrometric
measurements}
\label{table:8}
\begin{center}
\begin{tabular}{lcr@{\,$\pm$\,}lr@{\,$\pm$\,}l}
\hline\hline
Object & proj. & \multicolumn{2}{c}{Change in    } & \multicolumn{2}{c}{Change     }  \\
       & sep.  & \multicolumn{2}{c}{separation  } & \multicolumn{2}{c}{in PA$^b$   }  \\
       & [AU]  & \multicolumn{2}{c}{[mas/yr]$^a$} & \multicolumn{2}{c}{[$\degr$/yr]}  \\
\hline
DI Cha A(BC)$^c$          & 755       &   0.61 & 3.37 &  0.03  & 0.06 \\
\ \ \ \ \ \ \ \ 
\ \ \ \, BC               & \ \ 10    &  -2.89 & 0.54 & -4.94  & 0.44 \\
\ \ \ \ \ \ \ \ 
\ \ \ \, AD               & \ \ 35    &   1.12 & 1.05 &  2.18  & 0.28 \\
Sz 22 AB                  & \ \ 85    &   2.47 & 0.56 & -0.07  & 0.09 \\
CHXR 32 A(BC)$^c$         & 401       &   0.30 & 2.23 & -0.01  & 0.07 \\
\ \ \ \ \ \ \ \ 
\ \ \ \ \ \ \ \ 
\ A(BC)$^d$               & 398       &   0.03 & 2.79 &  0.03  & 0.08 \\
\ \ \ \ \ \ \ \ 
\ \ \ \ \ \ \ \ 
\ BC                      & \ \ 14    &   9.69 & 3.73 & -2.58  & 0.87 \\
Cha H$\alpha$ 5 AB        & \ \ \ \ 8 &  -0.75 & 1.14 &  7.84  & 2.97 \\
\ \ \ \ \ \ \ \ \ \ \
\ \ \ \ \, (AB)cc1$^e$    & (232)     &  -10.6 & 1.99 & -0.41  & 0.10 \\
\hline
\end{tabular}
\end{center}
\textbf{Notes.} $^{(a)}$ Using a nominal pixel scale of 0.01324 arcsec/pixel to convert from pixel to mas $^{(b)}$ PA is measured 
from N over E to S. 
$^{(c)}$ See $^{d}$ in Table \ref{table:5}. $^{(d)}$ See $^{e}$ in Table \ref{table:5}. $^{(e)}$ See $^{f}$ in Table \ref{table:5}.
\end{table}

\onltab{9}{
\begin{table*}
\caption{Measured brightness differences and mean apparent magnitudes}
\label{table:9}
\begin{center} 
\begin{tabular}{llccc|cccc}
\hline\hline
Object & Epoch & J-band & H-band & Ks-band & Ob-    & J-band & H-band & Ks-band \\
       &       & [mag]  & [mag]  & [mag]   & ject   & [mag]  & [mag]  & [mag]   \\
\hline   
DI Cha A(BC)        & 17 Feb 2006 &                     && \, \, 4.136 $\pm$ 0.010 & A   & \ \ 7.841 && \ \ 6.239 \\
                    & 19 Feb 2008 & 4.232 $\pm$ 0.018   && \, \, 4.347 $\pm$ 0.033 & BC  & 12.073    && 10.480    \\ 
                    & 20 Feb 2009$^b$ &                 && $>$ 4.104 $\pm$ 0.010   & B   & 12.603    && 11.077    \\
\ \ \ \ \ \ \ \ 
\ \ \ \, AB         & 17 Feb 2006 &                     && \, \, 4.715 $\pm$ 0.007 & C   & 12.749    && 11.087    \\
                    & 19 Feb 2008 & 4.770 $\pm$ 0.013   && \, \, 4.980 $\pm$ 0.035 & D   & \ 12.938$^a$&& 11.468  \\ 
                    & 20 Feb 2009$^b$ &                 && $>$ 4.737 $\pm$ 0.017   &     &           &&           \\
\ \ \ \ \ \ \ \ 
\ \ \ \, AC         & 17 Feb 2006 &                     && \, \, 4.741 $\pm$ 0.007 &     &           &&           \\
                    & 19 Feb 2008 & 4.919 $\pm$ 0.052   && \, \, 4.974 $\pm$ 0.029 &     &           &&           \\ 
                    & 20 Feb 2009$^b$ &                 && $>$ 4.784 $\pm$ 0.019   &     &           &&           \\
\ \ \ \ \ \ \ \ 
\ \ \ \, BC         & 17 Feb 2006 &                     && \, \, 0.026 $\pm$ 0.004 &     &           &&           \\
                    & 19 Feb 2008 & 0.147 $\pm$ 0.051   && \ \ -0.006 $\pm$ 0.013  &     &           &&           \\ 
                    & 20 Feb 2009 &                     && \, \, 0.046 $\pm$ 0.032 &     &           &&           \\
\ \ \ \ \ \ \ \ 
\ \ \ \, AD         & 17 Feb 2006 &                     && \, \, 5.298 $\pm$ 0.020 &     &           &&           \\
                    & 19 Feb 2008 & 5.097$^a$ $\pm$ 0.019&& \, \, 5.186 $\pm$ 0.016&     &           &&           \\ 
                    & 20 Feb 2009$^b$ &                 && $>$ 4.648 $\pm$ 0.006   &     &           &&           \\
Sz 22 AB            & 17 Feb 2006 &                     && \, \, 3.165 $\pm$ 0.006 & A   & \ \ 9.639 && \ \ 6.877 \\
                    & 19 Feb 2008 & 2.179 $\pm$ 0.007   && \, \, 3.654 $\pm$ 0.014 & B   & 11.818    && 10.286    \\
CHXR 32 A(BC)$^d$   & 25 Mar 2006$^c$&& 0.846 $\pm$ 0.008 &                        & A   && 7.599     & \ \ 6.350 \\
                    & 20 Feb 2008 &                     && \, \, 1.940 $\pm$ 0.002 & BC  && 8.445     & \ \ 8.290 \\
\ \ \ \ \ \ \ \ 
\ \ \ \ \ \ \ \ \ AB & 25 Mar 2006$^c$&                 && \, \, 2.150 $\pm$ 0.058 & B   &           && \ \ 8.527 \\
                    & 20 Feb 2009 &                     && \, \, 2.177 $\pm$ 0.005 & C   &           && \ \ 9.843 \\
\ \ \ \ \ \ \ \ 
\ \ \ \ \ \ \ \ \ AC & 25 Mar 2006$^c$&                 && \, \, 3.534 $\pm$ 0.182 &     &           &&           \\
                    & 20 Feb 2009 &                     && \, \, 3.493 $\pm$ 0.006 &     &           &&           \\
\ \ \ \ \ \ \ \ 
\ \ \ \ \ \ \ \ \ BC & 25 Mar 2006$^c$&                 && \, \, 1.384 $\pm$ 0.238 &     &           &&           \\
                    & 20 Feb 2009 &                     && \, \, 1.317 $\pm$ 0.004 &     &           &&           \\
Cha H$\alpha$ 5 AB  & 16 Feb 2006 &                     && \, \, 0.036 $\pm$ 0.146 & A   &           && 11.439    \\
                    & 19 Feb 2009 &                     && \, \, 0.153 $\pm$ 0.093 & B   &           && 11.534    \\
\ \ \ \ \ \ \ \ 
\ \ \ \ \ \ \ \, (AB)cc1$^d$ & 20 Feb 2008 &            && \, \, 5.234 $\pm$ 0.056 & cc1 &           && 15.945    \\
\ \ \ \ \ \ \ \ 
\ \ \ \ \ \ \ \, Acc1 & 16 Feb 2006 &                   && \, \, 4,471 $\pm$ 0.070 &     &           &&           \\
                    & 19 Feb 2009 &                     && \, \, 4.469 $\pm$ 0.044 &     &           &&           \\
\ \ \ \ \ \ \ \ 
\ \ \ \ \ \ \ \, Acc2 & 16 Feb 2006 &                   && \, \, 4,458 $\pm$ 0.005 &     &           &&           \\
                    & 19 Feb 2009 &                     && \, \, 4.397 $\pm$ 0.021 &     &           &&           \\
\hline
\end{tabular}
\end{center}
\textbf{Notes.} Mean apparent magnitudes based on combined brightness measurements of 2MASS \citep{2006AJ....131.1163S}.
$^{(a)}$ Peak to peak brightness difference given. PSF photometry results systematic too low after bad pixel correction in PSF, as bad pixel present despite jittering (only 5 images).
$^{(b)}$ primary (component A) saturated within this epoch.
$^{(c)}$ Rereduced, see also \citet{2008ApJ...683..844L}.
$^{(d)}$ PSF photometry does not work in the case of non-resolved (by fitting) double objects, aperture photometry used, see text for further information.
\end{table*}
}

\section{Conclusions}
\label{Section4}

The search for stellar and sub-stellar companions of young low-mass stars is one of our main aims, where we apply high-precision astrometry in the JHKs
band passes at different
epochs over several years. With the VLT NACO instrument we can, this way, resolve binary separations of 0.07'', corresponding to projected linear separations of 6 -- 10 AU between the closest
components (depending on the adopted distance to the Cha star-forming regions). But even below this limit, we found the unresolved binary Cha H$\alpha$ 5 at a separation of less than
0.05'' due to its elongated PSF, which could be analysed using the PSF fitting method with dedicated software. This limit connects our astrometric method of multiplicity detection to that of
the orbit determination via radial velocity measurements, filling in the still present lack of detecting projected binary separations in the range 3 -- 10 AU in 
very young nearby star-forming regions at intermediate distances of about 150 pc \citep{2008A&A...492..545J} or as the Chamaeleon I complex at 165\,$\pm$\,30 pc (see \citet{2008A&A...491..311S}
for a discussion).

The VLT observations have a wide range of coverage in brightness, down to Ks$\sim$ 20 mag, enabling the detection of sub-stellar companions as brown dwarfs and planets.
In the Cha regions, however, such objects seem to be rather scarce. Only one radial velocity example named Cha H$\alpha$ 8 \citep{2007ApJ...666L.113J} and two
examples from imaging, CHXR 73 \citep{2006ApJ...649..894L} and CT Cha \citep{2008A&A...491..311S}, as well as two other candidates have been identified so far,
Cha H$\alpha$ 2 \citep{2008A&A...484..413S} and T Cha \citep{2011A&A...528L...7H}, the latter additionally using the
sparse aperture masking technique. Other rejected candidates found by us, as well as the corresponding sub-stellar statistics, will
be presented in a forthcoming paper.
In contrast, stellar companions are very frequent, and in Paper I we were able to confirm the physical reality of thirteen binaries and three triple systems. The present paper presents four
additional cases, two binaries, one triple, and one quadruple system. In each of these four targets we were able to detect a hitherto unknown companion and
confirm that these objects are co-moving and showing some kind of orbital motion with high probability.

While all four newly found stellar companions in this publication are part of the Cha I star-forming region\footnote{In the inner 7\,x\,14 arcmin part of the Cha I complex centred
on HD 97048, as presented e.g. in an official poster of ESO created using data from the instrument FORS at
http://www.eso.org/public/images/eso9921c/}, each object is individually of special interest. DI Cha is to our best knowledge the first directly resolved quadruple star system in Chamaleon,
for which in addition orbital motion of both hierarchical pairs could be derived and is
presented in Tables \ref{table:7} \& \ref{table:8}. Around the star Sz 22 we could identify a stellar co-moving companion, as well as a prominent disc/reflection nebula of
$\gtrsim$ 2.5 arcsec size, corresponding to $\gtrsim$ 410 AU. Since a temperature of about 1300~K, needed to be visible thermically in the Ks band, is
very unlikely despite its youth, the disc/nebula must be seen in reflection. Since the primary, as a K7 star,
might be too faint in the optical for such a strong reflective luminosity of the disc/nebula, we assume that the disc/nebula is
illuminated by \object{HD 97048}, an A0 star at $\sim$ 37 arcsec projected separation, corresponding to $\sim$ 5900 AU at the distance of HD 97048 of 158\,$\pm$\,16 pc \citep{2007A&A...474..653V},
which is itself surrounded by a large directly detected disc of $\sim$ 700 AU \citep{2007AJ....133.2122D} found using the Hubble Space Telescope.

CHXR 32, often also called Glass I, was known to be a
star with infrared companion. While the nature of the infrared companion can be explained by a G-star weakened in the optical by an edge-on disc, \citet{2013ApJ...764..127K} find difficulties
fitting the spectrum of CHXR 32 B (called Glass Ia there). They can fit an X-Shooter spectrum by including a mid K-type stellar source, as
well as an early M dwarf, which is needed to fit the CO overtone absorption feature at 2300 nm. This is fully consistent with our resolved triple nature of CHXR 32, although
at the edge of our allowed spectral classification and in good agreement with the apparent magnitudes in Table \ref{table:9}. A spectral type of K4, as given by
\citet{1988A&A...207...46C}, is slightly outside the spectral class range we found for CHXR 32 B,
but might be influenced by the previously unknown multiplicity. A classification as K6 in \citet{2004ApJ...602..816L} is just within the spectral range determined by us.
Finally, Cha H$\alpha$ 5 is the closest stellar companion found in this survey, but is not even fully resolved. The
components are still brown dwarf candidates as discussed for the fainter component of the very similar case of Cha H$\alpha$ 2 \citep{2008A&A...484..413S}, while this new binary has a much lower
separation, hence an orbital period of about 46 years, which will very soon allow curvature to be detected in the orbital motion.

We cannot exclude DI Cha D and Cha H$\alpha$ 5 A and B being brown dwarfs,
because their 1 $\sigma$ error minimum masses, which are 67, 68
and 65 M$_{\mathrm{Jup}}$ using their brightnesses
(Table \ref{table:9}) and COND evolutionary models \citep{2003A&A...402..701B} at 1 Myr age and minimum distance, respectively, are below the mass of the least massive stars of
75 M$_{\mathrm{Jup}}$ \citep{2000ARA&A..38..485B}.
However, at the nominal distance of Cha I of 165 pc all sub-stellar candidate objects have masses
$\geq$\,78 M$_{\mathrm{Jup}}$ even if we assume very young ages of 1 Myr. Both components of Cha H$\alpha$ 5
have masses of about 0.12 M$_{\sun}$ each ($\geq$\,83 M$_{\mathrm{Jup}}$ at 1 $\sigma$ error) as calculated using Kepler's third law and
assuming a strictly pole-on orbit and the measured orbital motion of the Cha H$\alpha$ 5 system within our two epochs (Tables \ref{table:7} \&
\ref{table:8}). Comparable masses of about 0.11 and 0.124 M$_{\sun}$ were found for the similar binary system Cha H$\alpha$ 2  on the basis of photometric
and spectroscopic observations \citep{2008A&A...484..413S}. If the total mass of the system was found to be $\leq$\,0.2 M$_{\sun}$, Cha H$\alpha$ 5 would
be a new member for the Very Low Mass Binaries Archive
\citep{2007lyot.confR..45S}, currently being composed from 99 systems, e.\,g.~2MASS J11011926-7732383 AB \citep{2004ApJ...614..398L} and Cha H$\alpha$ 8
AB \citep{2007ApJ...666L.113J} as members of Cha I. DI Cha D, on the other hand, possesses a brightness difference (Table \ref{table:9}) comparable to the recently found brown
dwarf companion PZ Tel B \citep{2010A&A...523L...1M,2010ApJ...720L..82B} at even smaller angular separation to its primary, while most likely having a
stellar nature due to its primary's higher mass.


\begin{acknowledgements}
We would like to thank the ESO Paranal Team, the ESO User Support department, and all the other very helpful ESO services as well as 
the anonymous referee, the 2 editors, C. Bertout and  T. Forveille, and the language editor J. Adams, for helpful comments.
Moreover, we thank D. Haase for providing his speckle pattern detection program `ringscale'.

TOBS acknowledges support from the Evangelisches Studienwerk e.V. Villigst.
NV acknowledges support by the Comit\'e Mixto ESO-Gobierno de Chile, as well as by the Gemini-CONICYT fund 32090027.
TOBS, RN \& TR would like to acknowledge support from the German National Science 
Foundation (Deutsche Forschungsgemeinschaft, DFG) in grant NE 515/30-1,
AB \& RN would like to acknowledge financial support from projects NE 515/13-1 and NE 515/13-2,
and TR \& RN would further like to acknowledge financial support under the project numbers NE 515/23-1 and NE 515/36-1.

HST data were obtained from the data archive at the Space Telescope Institute, which is operated by the association of
Universities for Research in Astronomy, Inc. under NASA contract NAS 5-26555.
This publication makes use of data products from the Two Micron All Sky Survey, which is a joint project of the University of
Massachusetts and the Infrared Processing and Analysis Center/California Institute of Technology, funded by the National 
Aeronautics and Space Administration and the National Science Foundation.
This research made use of the VizieR catalogue access tool and the Simbad database, both operated at the Observatoire 
de Strasbourg.
This reasearch makes use of the Hipparcos Catalogue, the primary result of the Hipparcos space astrometry mission, undertaken
by the European Space Agency.
This research made use of NASA's Astrophysics Data System Bibliographic Services.
This publication made use of the Very-Low-Mass Binaries Archive housed at http://www.vlmbinaries.org and maintained by Nick Siegler, Chris Gelino, and Adam Burgasser.
\end{acknowledgements}

\bibliographystyle{aa}

\begin{thebibliography}{70}
\expandafter\ifx\csname natexlab\endcsname\relax\def\natexlab#1{#1}\fi

\bibitem[{{Bally} {et~al.}(2006){Bally}, {Walawender}, {Luhman}, \&
  {Fazio}}]{2006AJ....132.1923B}
{Bally}, J., {Walawender}, J., {Luhman}, K.~L., \& {Fazio}, G. 2006, \aj, 132,
  1923

\bibitem[{{Banse} {et~al.}(1992){Banse}, {Grosbol}, \&
  {Baade}}]{1992ASPC...25..120B}
{Banse}, K., {Grosbol}, P., \& {Baade}, D. 1992, in Astronomical Society of the
  Pacific Conference Series, Vol.~25, Astronomical Data Analysis Software and
  Systems I, ed. D.~M. {Worrall}, C.~{Biemesderfer}, \& J.~{Barnes}, 120

\bibitem[{{Baraffe} {et~al.}(1998){Baraffe}, {Chabrier}, {Allard}, \&
  {Hauschildt}}]{1998A&A...337..403B}
{Baraffe}, I., {Chabrier}, G., {Allard}, F., \& {Hauschildt}, P.~H. 1998, \aap,
  337, 403

\bibitem[{{Baraffe} {et~al.}(2002){Baraffe}, {Chabrier}, {Allard}, \&
  {Hauschildt}}]{2002A&A...382..563B}
{Baraffe}, I., {Chabrier}, G., {Allard}, F., \& {Hauschildt}, P.~H. 2002, \aap,
  382, 563

\bibitem[{{Baraffe} {et~al.}(2003){Baraffe}, {Chabrier}, {Barman}, {Allard}, \&
  {Hauschildt}}]{2003A&A...402..701B}
{Baraffe}, I., {Chabrier}, G., {Barman}, T.~S., {Allard}, F., \& {Hauschildt},
  P.~H. 2003, \aap, 402, 701

\bibitem[{{Basri}(2000)}]{2000ARA&A..38..485B}
{Basri}, G. 2000, \araa, 38, 485

\bibitem[{{Bertout} {et~al.}(1999){Bertout}, {Robichon}, \&
  {Arenou}}]{1999A&A...352..574B}
{Bertout}, C., {Robichon}, N., \& {Arenou}, F. 1999, \aap, 352, 574

\bibitem[{{Biller} {et~al.}(2010){Biller}, {Liu}, {Wahhaj}, {Nielsen}, {Close},
  {Dupuy}, {Hayward}, {Burrows}, {Chun}, {Ftaclas}, {Clarke}, {Hartung},
  {Males}, {Reid}, {Shkolnik}, {Skemer}, {Tecza}, {Thatte}, {Alencar},
  {Artymowicz}, {Boss}, {de Gouveia Dal Pino}, {Gregorio-Hetem}, {Ida},
  {Kuchner}, {Lin}, \& {Toomey}}]{2010ApJ...720L..82B}
{Biller}, B.~A., {Liu}, M.~C., {Wahhaj}, Z., {et~al.} 2010, \apjl, 720, L82

\bibitem[{{Cambresy} {et~al.}(1997){Cambresy}, {Epchtein}, {Copet}, {de Batz},
  {Kimeswenger}, {Le Bertre}, {Rouan}, \& {Tiphene}}]{1997A&A...324L...5C}
{Cambresy}, L., {Epchtein}, N., {Copet}, E., {et~al.} 1997, \aap, 324, L5

\bibitem[{{Carpenter} {et~al.}(2002){Carpenter}, {Hillenbrand}, {Skrutskie}, \&
  {Meyer}}]{2002AJ....124.1001C}
{Carpenter}, J.~M., {Hillenbrand}, L.~A., {Skrutskie}, M.~F., \& {Meyer}, M.~R.
  2002, \aj, 124, 1001

\bibitem[{{Chelli} {et~al.}(1988){Chelli}, {Cruz-Gonzalez}, {Zinnecker},
  {Carrasco}, \& {Perrier}}]{1988A&A...207...46C}
{Chelli}, A., {Cruz-Gonzalez}, I., {Zinnecker}, H., {Carrasco}, L., \&
  {Perrier}, C. 1988, \aap, 207, 46

\bibitem[{{Comer{\'o}n} {et~al.}(2000){Comer{\'o}n}, {Neuh{\"a}user}, \&
  {Kaas}}]{2000A&A...359..269C}
{Comer{\'o}n}, F., {Neuh{\"a}user}, R., \& {Kaas}, A.~A. 2000, \aap, 359, 269

\bibitem[{{Comer{\'o}n} {et~al.}(1999){Comer{\'o}n}, {Rieke}, \&
  {Neuh{\"a}user}}]{1999A&A...343..477C}
{Comer{\'o}n}, F., {Rieke}, G.~H., \& {Neuh{\"a}user}, R. 1999, \aap, 343, 477

\bibitem[{{Correia} {et~al.}(2006){Correia}, {Zinnecker}, {Ratzka}, \&
  {Sterzik}}]{2006A&A...459..909C}
{Correia}, S., {Zinnecker}, H., {Ratzka}, T., \& {Sterzik}, M.~F. 2006, \aap,
  459, 909

\bibitem[{{Cutri} {et~al.}(2003){Cutri}, {Skrutskie}, {van Dyk}, {Beichman},
  {Carpenter}, {Chester}, {Cambresy}, {Evans}, {Fowler}, {Gizis}, {Howard},
  {Huchra}, {Jarrett}, {Kopan}, {Kirkpatrick}, {Light}, {Marsh}, {McCallon},
  {Schneider}, {Stiening}, {Sykes}, {Weinberg}, {Wheaton}, {Wheelock}, \&
  {Zacarias}}]{2003tmc..book.....C}
{Cutri}, R.~M., {Skrutskie}, M.~F., {van Dyk}, S., {et~al.} 2003, {2MASS All
  Sky Catalog of point sources.}, ed. R.~M. {Cutri}, M.~F. {Skrutskie}, S.~{van
  Dyk}, C.~A. {Beichman}, J.~M. {Carpenter}, T.~{Chester}, L.~{Cambresy},
  T.~{Evans}, J.~{Fowler}, J.~{Gizis}, E.~{Howard}, J.~{Huchra}, T.~{Jarrett},
  E.~L. {Kopan}, J.~D. {Kirkpatrick}, R.~M. {Light}, K.~A. {Marsh},
  H.~{McCallon}, S.~{Schneider}, R.~{Stiening}, M.~{Sykes}, M.~{Weinberg},
  W.~A. {Wheaton}, S.~{Wheelock}, \& N.~{Zacarias}

\bibitem[{{Diolaiti} {et~al.}(2000){Diolaiti}, {Bendinelli}, {Bonaccini},
  {Close}, {Currie}, \& {Parmeggiani}}]{2000SPIE.4007..879D}
{Diolaiti}, E., {Bendinelli}, O., {Bonaccini}, D., {et~al.} 2000, in Presented
  at the Society of Photo-Optical Instrumentation Engineers (SPIE) Conference,
  Vol. 4007, Proc. SPIE Vol. 4007, p. 879-888, Adaptive Optical Systems
  Technology, Peter L. Wizinowich; Ed., ed. P.~L. {Wizinowich}, 879--888

\bibitem[{{Doering} {et~al.}(2007){Doering}, {Meixner}, {Holfeltz}, {Krist},
  {Ardila}, {Kamp}, {Clampin}, \& {Lubow}}]{2007AJ....133.2122D}
{Doering}, R.~L., {Meixner}, M., {Holfeltz}, S.~T., {et~al.} 2007, \aj, 133,
  2122

\bibitem[{{Ducourant} {et~al.}(2005){Ducourant}, {Teixeira}, {P{\'e}ri{\'e}},
  {Lecampion}, {Guibert}, \& {Sartori}}]{2005A&A...438..769D}
{Ducourant}, C., {Teixeira}, R., {P{\'e}ri{\'e}}, J.~P., {et~al.} 2005, \aap,
  438, 769

\bibitem[{{Feigelson} \& {Kriss}(1989)}]{1989ApJ...338..262F}
{Feigelson}, E.~D. \& {Kriss}, G.~A. 1989, \apj, 338, 262

\bibitem[{{Furlan} {et~al.}(2009){Furlan}, {Watson}, {McClure}, {Manoj},
  {Espaillat}, {D'Alessio}, {Calvet}, {Kim}, {Sargent}, {Forrest}, \&
  {Hartmann}}]{2009ApJ...703.1964F}
{Furlan}, E., {Watson}, D.~M., {McClure}, M.~K., {et~al.} 2009, \apj, 703, 1964

\bibitem[{{Ghez} {et~al.}(1997){Ghez}, {McCarthy}, {Patience}, \&
  {Beck}}]{1997ApJ...481..378G}
{Ghez}, A.~M., {McCarthy}, D.~W., {Patience}, J.~L., \& {Beck}, T.~L. 1997,
  \apj, 481, 378

\bibitem[{{G{\'o}mez} \& {Mardones}(2003)}]{2003AJ....125.2134G}
{G{\'o}mez}, M. \& {Mardones}, D. 2003, \aj, 125, 2134

\bibitem[{{Guenther} {et~al.}(2007){Guenther}, {Esposito}, {Mundt}, {Covino},
  {Alcal{\'a}}, {Cusano}, \& {Stecklum}}]{2007A&A...467.1147G}
{Guenther}, E.~W., {Esposito}, M., {Mundt}, R., {et~al.} 2007, \aap, 467, 1147

\bibitem[{{Haase}(2009)}]{HaaseD2009}
{Haase}, D. 2009, Report about Student Research Project, Univ. Jena

\bibitem[{{Haisch} {et~al.}(2004){Haisch}, {Greene}, {Barsony}, \&
  {Stahler}}]{2004AJ....127.1747H}
{Haisch}, Jr., K.~E., {Greene}, T.~P., {Barsony}, M., \& {Stahler}, S.~W. 2004,
  \aj, 127, 1747

\bibitem[{{Hambly} {et~al.}(2001){Hambly}, {MacGillivray}, {Read}, {Tritton},
  {Thomson}, {Kelly}, {Morgan}, {Smith}, {Driver}, {Williamson}, {Parker},
  {Hawkins}, {Williams}, \& {Lawrence}}]{2001MNRAS.326.1279H}
{Hambly}, N.~C., {MacGillivray}, H.~T., {Read}, M.~A., {et~al.} 2001, \mnras,
  326, 1279

\bibitem[{{Henize} \& {Mendoza}(1973)}]{1973ApJ...180..115H}
{Henize}, K.~G. \& {Mendoza}, E.~E. 1973, \apj, 180, 115

\bibitem[{{H{\o}g} {et~al.}(2000){H{\o}g}, {Fabricius}, {Makarov}, {Urban},
  {Corbin}, {Wycoff}, {Bastian}, {Schwekendiek}, \&
  {Wicenec}}]{2000A&A...355L..27H}
{H{\o}g}, E., {Fabricius}, C., {Makarov}, V.~V., {et~al.} 2000, \aap, 355, L27

\bibitem[{{Hu{\'e}lamo} {et~al.}(2011){Hu{\'e}lamo}, {Lacour}, {Tuthill},
  {Ireland}, {Kraus}, \& {Chauvin}}]{2011A&A...528L...7H}
{Hu{\'e}lamo}, N., {Lacour}, S., {Tuthill}, P., {et~al.} 2011, \aap, 528, L7

\bibitem[{{Joergens}(2008)}]{2008A&A...492..545J}
{Joergens}, V. 2008, \aap, 492, 545

\bibitem[{{Joergens} \& {M{\"u}ller}(2007)}]{2007ApJ...666L.113J}
{Joergens}, V. \& {M{\"u}ller}, A. 2007, \apjl, 666, L113

\bibitem[{{Kainulainen} {et~al.}(2006){Kainulainen}, {Lehtinen}, \&
  {Harju}}]{2006A&A...447..597K}
{Kainulainen}, J., {Lehtinen}, K., \& {Harju}, J. 2006, \aap, 447, 597

\bibitem[{{Kenyon} \& {Hartmann}(1995)}]{1995ApJS..101..117K}
{Kenyon}, S.~J. \& {Hartmann}, L. 1995, \apjs, 101, 117

\bibitem[{{Kraus} \& {Hillenbrand}(2007)}]{2007ApJ...662..413K}
{Kraus}, A.~L. \& {Hillenbrand}, L.~A. 2007, \apj, 662, 413

\bibitem[{{Kruger} {et~al.}(2013){Kruger}, {Richter}, {Carr}, {Najita},
  {Moerchen}, {Doppmann}, \& {Seifahrt}}]{2013ApJ...764..127K}
{Kruger}, A.~J., {Richter}, M.~J., {Carr}, J.~S., {et~al.} 2013, \apj, 764, 127

\bibitem[{{Lafreni{\`e}re} {et~al.}(2008){Lafreni{\`e}re}, {Jayawardhana},
  {Brandeker}, {Ahmic}, \& {van Kerkwijk}}]{2008ApJ...683..844L}
{Lafreni{\`e}re}, D., {Jayawardhana}, R., {Brandeker}, A., {Ahmic}, M., \& {van
  Kerkwijk}, M.~H. 2008, \apj, 683, 844

\bibitem[{{Lenzen} {et~al.}(2003){Lenzen}, {Hartung}, {Brandner}, {Finger},
  {Hubin}, {Lacombe}, {Lagrange}, {Lehnert}, {Moorwood}, \&
  {Mouillet}}]{2003SPIE.4841..944L}
{Lenzen}, R., {Hartung}, M., {Brandner}, W., {et~al.} 2003, in Presented at the
  Society of Photo-Optical Instrumentation Engineers (SPIE) Conference, Vol.
  4841, Instrument Design and Performance for Optical/Infrared Ground-based
  Telescopes. Edited by Iye, Masanori; Moorwood, Alan F. M. Proceedings of the
  SPIE, Volume 4841, pp. 944-952 (2003)., ed. M.~{Iye} \& A.~F.~M. {Moorwood},
  944--952

\bibitem[{{Lommen} {et~al.}(2007){Lommen}, {Wright}, {Maddison},
  {J{\o}rgensen}, {Bourke}, {van Dishoeck}, {Hughes}, {Wilner}, {Burton}, \&
  {van Langevelde}}]{2007A&A...462..211L}
{Lommen}, D., {Wright}, C.~M., {Maddison}, S.~T., {et~al.} 2007, \aap, 462, 211

\bibitem[{{L{\'o}pez Mart{\'{\i}}} {et~al.}(2004){L{\'o}pez Mart{\'{\i}}},
  {Eisl{\"o}ffel}, {Scholz}, \& {Mundt}}]{2004A&A...416..555L}
{L{\'o}pez Mart{\'{\i}}}, B., {Eisl{\"o}ffel}, J., {Scholz}, A., \& {Mundt}, R.
  2004, \aap, 416, 555

\bibitem[{{Luhman}(2004{\natexlab{a}})}]{2004ApJ...602..816L}
{Luhman}, K.~L. 2004{\natexlab{a}}, \apj, 602, 816

\bibitem[{{Luhman}(2004{\natexlab{b}})}]{2004ApJ...614..398L}
{Luhman}, K.~L. 2004{\natexlab{b}}, \apj, 614, 398

\bibitem[{{Luhman}(2007)}]{2007ApJS..173..104L}
{Luhman}, K.~L. 2007, \apjs, 173, 104

\bibitem[{{Luhman}(2008)}]{2008hsf2.book..169L}
{Luhman}, K.~L. 2008, {Chamaeleon}, ed. {Reipurth, B.}, 169--+

\bibitem[{{Luhman} {et~al.}(2008){Luhman}, {Allen}, {Allen}, {Gutermuth},
  {Hartmann}, {Mamajek}, {Megeath}, {Myers}, \& {Fazio}}]{2008ApJ...675.1375L}
{Luhman}, K.~L., {Allen}, L.~E., {Allen}, P.~R., {et~al.} 2008, \apj, 675, 1375

\bibitem[{{Luhman} {et~al.}(2003){Luhman}, {Stauffer}, {Muench}, {Rieke},
  {Lada}, {Bouvier}, \& {Lada}}]{2003ApJ...593.1093L}
{Luhman}, K.~L., {Stauffer}, J.~R., {Muench}, A.~A., {et~al.} 2003, \apj, 593,
  1093

\bibitem[{{Luhman} {et~al.}(2006){Luhman}, {Wilson}, {Brandner}, {Skrutskie},
  {Nelson}, {Smith}, {Peterson}, {Cushing}, \& {Young}}]{2006ApJ...649..894L}
{Luhman}, K.~L., {Wilson}, J.~C., {Brandner}, W., {et~al.} 2006, \apj, 649, 894

\bibitem[{{Manoj} {et~al.}(2011){Manoj}, {Kim}, {Furlan}, {McClure}, {Luhman},
  {Watson}, {Espaillat}, {Calvet}, {Najita}, {D'Alessio}, {Adame}, {Sargent},
  {Forrest}, {Bohac}, {Green}, \& {Arnold}}]{2011ApJS..193...11M}
{Manoj}, P., {Kim}, K.~H., {Furlan}, E., {et~al.} 2011, \apjs, 193, 11

\bibitem[{{Melo}(2003)}]{2003A&A...410..269M}
{Melo}, C.~H.~F. 2003, \aap, 410, 269

\bibitem[{{Mugrauer} {et~al.}(2010){Mugrauer}, {Vogt}, {Neuh{\"a}user}, \&
  {Schmidt}}]{2010A&A...523L...1M}
{Mugrauer}, M., {Vogt}, N., {Neuh{\"a}user}, R., \& {Schmidt}, T.~O.~B. 2010,
  \aap, 523, L1+

\bibitem[{{Neuh{\"a}user} {et~al.}(2002){Neuh{\"a}user}, {Brandner}, {Alves},
  {Joergens}, \& {Comer{\'o}n}}]{2002A&A...384..999N}
{Neuh{\"a}user}, R., {Brandner}, W., {Alves}, J., {Joergens}, V., \&
  {Comer{\'o}n}, F. 2002, \aap, 384, 999

\bibitem[{{Neuh{\"a}user} \& {Comeron}(1998)}]{1998Sci...282...83N}
{Neuh{\"a}user}, R. \& {Comeron}, F. 1998, Science, 282, 83

\bibitem[{{Neuh{\"a}user} {et~al.}(2003){Neuh{\"a}user}, {Guenther}, \&
  {Brandner}}]{2003IAUS..211..309N}
{Neuh{\"a}user}, R., {Guenther}, E., \& {Brandner}, W. 2003, in IAU Symposium,
  Vol. 211, Brown Dwarfs, ed. {E.~Mart{\'{\i}}n}, 309

\bibitem[{{Nguyen} {et~al.}(2012){Nguyen}, {Brandeker}, {van Kerkwijk}, \&
  {Jayawardhana}}]{2012ApJ...745..119N}
{Nguyen}, D.~C., {Brandeker}, A., {van Kerkwijk}, M.~H., \& {Jayawardhana}, R.
  2012, \apj, 745, 119

\bibitem[{{Padgett} \& {Stapelfeldt}(2001)}]{2001AAS...199.6009P}
{Padgett}, D.~L. \& {Stapelfeldt}, K.~R. 2001, in Bulletin of the American
  Astronomical Society, Vol.~33, Bulletin of the American Astronomical Society,
  1395--+

\bibitem[{{Perryman} {et~al.}(1997){Perryman}, {Lindegren}, {Kovalevsky},
  {Hoeg}, {Bastian}, {Bernacca}, {Cr{\'e}z{\'e}}, {Donati}, {Grenon}, {van
  Leeuwen}, {van der Marel}, {Mignard}, {Murray}, {Le Poole}, {Schrijver},
  {Turon}, {Arenou}, {Froeschl{\'e}}, \& {Petersen}}]{1997A&A...323L..49P}
{Perryman}, M.~A.~C., {Lindegren}, L., {Kovalevsky}, J., {et~al.} 1997, \aap,
  323, L49

\bibitem[{{Reipurth} \& {Zinnecker}(1993)}]{1993A&A...278...81R}
{Reipurth}, B. \& {Zinnecker}, H. 1993, \aap, 278, 81

\bibitem[{{Rieke} \& {Lebofsky}(1985)}]{1985ApJ...288..618R}
{Rieke}, G.~H. \& {Lebofsky}, M.~J. 1985, \apj, 288, 618

\bibitem[{{R{\"o}ser} {et~al.}(2008){R{\"o}ser}, {Schilbach}, {Schwan},
  {Kharchenko}, {Piskunov}, \& {Scholz}}]{2008A&A...488..401R}
{R{\"o}ser}, S., {Schilbach}, E., {Schwan}, H., {et~al.} 2008, \aap, 488, 401

\bibitem[{{Rousset} {et~al.}(2003){Rousset}, {Lacombe}, {Puget}, {Hubin},
  {Gendron}, {Fusco}, {Arsenault}, {Charton}, {Feautrier}, {Gigan}, {Kern},
  {Lagrange}, {Madec}, {Mouillet}, {Rabaud}, {Rabou}, {Stadler}, \&
  {Zins}}]{2003SPIE.4839..140R}
{Rousset}, G., {Lacombe}, F., {Puget}, P., {et~al.} 2003, in Presented at the
  Society of Photo-Optical Instrumentation Engineers (SPIE) Conference, Vol.
  4839, Adaptive Optical System Technologies II. Edited by Wizinowich, Peter
  L.; Bonaccini, Domenico. Proceedings of the SPIE, Volume 4839, pp. 140-149
  (2003)., ed. P.~L. {Wizinowich} \& D.~{Bonaccini}, 140--149

\bibitem[{{Schmidt} {et~al.}(2008{\natexlab{a}}){Schmidt}, {Neuh{\"a}user},
  {Seifahrt}, {Vogt}, {Bedalov}, {Helling}, {Witte}, \&
  {Hauschildt}}]{2008A&A...491..311S}
{Schmidt}, T.~O.~B., {Neuh{\"a}user}, R., {Seifahrt}, A., {et~al.}
  2008{\natexlab{a}}, \aap, 491, 311

\bibitem[{{Schmidt} {et~al.}(2008{\natexlab{b}}){Schmidt}, {Neuh{\"a}user},
  {Vogt}, {Seifahrt}, {Roell}, \& {Bedalov}}]{2008A&A...484..413S}
{Schmidt}, T.~O.~B., {Neuh{\"a}user}, R., {Vogt}, N., {et~al.}
  2008{\natexlab{b}}, \aap, 484, 413

\bibitem[{{Siegler}(2007)}]{2007lyot.confR..45S}
{Siegler}, N. 2007, in In the Spirit of Bernard Lyot: The Direct Detection of
  Planets and Circumstellar Disks in the 21st Century, 45

\bibitem[{{Skrutskie} {et~al.}(2006){Skrutskie}, {Cutri}, {Stiening},
  {Weinberg}, {Schneider}, {Carpenter}, {Beichman}, {Capps}, {Chester},
  {Elias}, {Huchra}, {Liebert}, {Lonsdale}, {Monet}, {Price}, {Seitzer},
  {Jarrett}, {Kirkpatrick}, {Gizis}, {Howard}, {Evans}, {Fowler}, {Fullmer},
  {Hurt}, {Light}, {Kopan}, {Marsh}, {McCallon}, {Tam}, {Van Dyk}, \&
  {Wheelock}}]{2006AJ....131.1163S}
{Skrutskie}, M.~F., {Cutri}, R.~M., {Stiening}, R., {et~al.} 2006, \aj, 131,
  1163

\bibitem[{{Stapelfeldt}(2001)}]{2001ASPC..231..620S}
{Stapelfeldt}, K. 2001, in Astronomical Society of the Pacific Conference
  Series, Vol. 231, Tetons 4: Galactic Structure, Stars and the Interstellar
  Medium, ed. C.~E. {Woodward}, M.~D. {Bicay}, \& J.~M. {Shull}, 620--+

\bibitem[{{Tetzlaff} {et~al.}(2011){Tetzlaff}, {Neuh{\"a}user}, \&
  {Hohle}}]{2011MNRAS.410..190T}
{Tetzlaff}, N., {Neuh{\"a}user}, R., \& {Hohle}, M.~M. 2011, \mnras, 410, 190

\bibitem[{{van Leeuwen}(2007)}]{2007A&A...474..653V}
{van Leeuwen}, F. 2007, \aap, 474, 653

\bibitem[{{Vogt} {et~al.}(2012){Vogt}, {Schmidt}, {Neuh{\"a}user}, {Bedalov},
  {Roell}, {Seifahrt}, \& {Mugrauer}}]{2012A&A...546A..63V}
{Vogt}, N., {Schmidt}, T.~O.~B., {Neuh{\"a}user}, R., {et~al.} 2012, \aap, 546,
  A63 (Paper I)

\bibitem[{{Wenger} {et~al.}(2007){Wenger}, {Oberto}, {Bonnarel}, {Brouty},
  {Bruneau}, {Brunet}, {Cambresy}, {Dubois}, {Eisele}, {Fernique}, {Genova},
  {Lalo{\"e}}, {Lesteven}, {Loup}, {Ochsenbein}, {Vannier}, {Vollmer},
  {Vonflie}, {Wagner}, {Woelfel}, {Borde}, {Beyneix}, {Chassagnard},
  {Jasniewicz}, \& {Davoust}}]{2007ASPC..377..197W}
{Wenger}, M., {Oberto}, A., {Bonnarel}, F., {et~al.} 2007, in Astronomical
  Society of the Pacific Conference Series, Vol. 377, Library and Information
  Services in Astronomy V, ed. {S.~Ricketts, C.~Birdie, \& E.~Isaksson}, 197--+

\bibitem[{{Zacharias} {et~al.}(2010){Zacharias}, {Finch}, {Girard}, {Hambly},
  {Wycoff}, {Zacharias}, {Castillo}, {Corbin}, {DiVittorio}, {Dutta}, {Gaume},
  {Gauss}, {Germain}, {Hall}, {Hartkopf}, {Hsu}, {Holdenried}, {Makarov},
  {Martinez}, {Mason}, {Monet}, {Rafferty}, {Rhodes}, {Siemers}, {Smith},
  {Tilleman}, {Urban}, {Wieder}, {Winter}, \& {Young}}]{2010AJ....139.2184Z}
{Zacharias}, N., {Finch}, C., {Girard}, T., {et~al.} 2010, \aj, 139, 2184

\bibitem[{{Zacharias} {et~al.}(2004){Zacharias}, {Urban}, {Zacharias},
  {Wycoff}, {Hall}, {Monet}, \& {Rafferty}}]{2004AJ....127.3043Z}
{Zacharias}, N., {Urban}, S.~E., {Zacharias}, M.~I., {et~al.} 2004, \aj, 127,
  3043

\end{thebibliography}

\listofobjects

\end{document}